\documentclass[12pt]{article}
\usepackage{amsmath,epsfig,array,lscape,hyperref}
\numberwithin{equation}{section}

\newcommand{\nc}{\newcommand} \nc{\rnc}{\renewcommand}
\rnc{\cases}[1]{\left\{\begin{array}{ll} #1 \end{array}\right.}
\rnc{\title}[1]{{\Large\bf\mbox{}\\\medskip#1\bigskip\medskip\\ }}
\rnc{\author}[1]{{\large #1\smallskip\\ }} \nc{\address}[1]{{\em
#1\medskip\\ }} \nc{\e}[1]{{\em #1\/}} \nc{\comment}[1]{}
\nc{\itm}[2]{\\\noindent$\bullet$\ \ \e{#1}.  \ #2}
\nc{\ru}[1]{\rule[-#1ex]{0ex}{#1ex}} \rnc{\baselinestretch}{1.1}
\rnc{\arraystretch}{0.91} \nc{\Id}{\mathrm{Id}}
\def\swnedots{\mathinner{\mkern1mu\raise1pt\vbox{\kern7pt\hbox{.}}\mkern2mu
\raise4pt\hbox{.}\mkern2mu\raise7pt\hbox{.}\mkern1mu}}
\nc{\be}{\begin{equation}}
\nc{\ee}{\end{equation}}
\nc{\ba}[1]{\begin{array}{@{}#1@{}}}
\nc{\ea}{\end{array}}
\nc{\disp}{\displaystyle}
\nc{\ade}{\mbox{$A$-$D$-$E$}}
\nc{\calA}{{\cal A}}
\nc{\calH}{{\cal H}}
\nc{\calB}{{\cal B}}
\nc{\calC}{{\cal C}}
\nc{\calE}{{\cal E}}
\nc{\calI}{{\cal I}}
\nc{\calM}{{\cal M}}
\nc{\calN}{{\cal N}}
\nc{\calS}{{\cal S}}
\nc{\calV}{{\cal V}}

\def\half {\mbox{$\textstyle \frac12$}}

\def\non{\nonumber}

\def\({\left(}
\def\){\right)}
\nc{\ds}{\displaystyle}

\def\topped#1#2{\genfrac{}{}{0pt}{3}{#1}{#2}}

\rnc{\Re}{\text{Re~}}
\rnc{\Im}{\text{Im~}}

\font\tenmsb=msbm10 scaled \magstep 2
\font\sevenmsb=msbm10 scaled \magstep 1
\font\fivemsb=msbm10
\newfam\msbfam
\textfont\msbfam=\sevenmsb
\scriptfont\msbfam=\sevenmsb
\scriptscriptfont\msbfam=\fivemsb
\def\Bbb#1{{\fam\msbfam\relax#1}}
\newfam\smmsbfam
\textfont\smmsbfam=\sevenmsb
\def\smBbb#1{{\fam\smmsbfam\relax #1}}
\newfam\Bmsbfam
\textfont\Bmsbfam=\tenmsb
\def\BBbb#1{{\fam\Bmsbfam\relax #1}}

\font\bigtenmsb=msbm10 scaled \magstep 3
\font\bigsevenmsb=msbm7 scaled \magstep 3
\font\bigfivemsb=msbm5 scaled \magstep 3
\newfam\bigmsbfam
\textfont\bigmsbfam=\bigtenmsb
\scriptfont\bigmsbfam=\bigsevenmsb
\scriptscriptfont\bigmsbfam=\bigfivemsb
\def\Bigbb#1{{\fam\bigmsbfam\relax#1}}

\nc{\W}[5]{W\!\left(\,\begin{array}{@{}cc|@{\:}}#4&#3\\
#1&#2\end{array}\;#5\right)}

\nc{\B}[7]{B^{#1}\!\!\left(\left.\!#4\ba{@{\;\,}c@{\;\,}c@{\:}}#5&#6\\
#2&#3\ea\right|#7\right)}
\rnc{\ss}{\scriptstyle}
\nc{\sss}{\scriptscriptstyle}
\nc{\hf}{{\sss1\!/\!2}}
\nc{\p}[2]{\makebox(0,0)[#1]{$#2$}}
\nc{\pp}[2]{\makebox(0,0)[#1]{$\ss#2$}}
\nc{\textB}[6]{\begin{picture}(#1,#2)\put(#3,#4){\p{#5}{\ds#6}}\end{picture}}
\rnc{\vec}[1]{\mbox{\boldmath$#1$}}

\begin{document}
\hfill\today
\begin{center}
\title{Integrable and Conformal Boundary Conditions\\
for ${\Bigbb Z}_k$ Parafermions on a Cylinder}
\author{Christian Mercat and Paul A. Pearce}
\address{Department of Mathematics and Statistics\\
University of Melbourne\\Parkville, Victoria 3010, Australia}
\medskip
\begin{abstract}
We study integrable and conformal boundary conditions for
$\hat{s\ell}(2)$ ${\Bbb Z}_k$ parafermions on a cylinder. These
conformal field theories are realized as the continuum scaling limit
of critical
$A$-$D$-$E$ lattice models with negative spectral parameter. The conformal
boundary conditions labelled by $(a,m)\in(G,{\Bbb Z}_{2k})$ are identified with
associated integrable lattice boundary conditions labelled by
$(r,a)\in(A_{g-2},G)$ where $g$ is the Coxeter number of the $A$-$D$-$E$ graph
$G$. We obtain analytically the boundary free energies, present general
expressions for the parafermion cylinder partition functions and
confirm these results by numerical
calculations.
\end{abstract}
\end{center}

\section{Introduction}

There has been much recent
progress~\cite{BePZ,BPPZ98,BPPZ00,BehrendP,PZ00,PZ01,PSS96} on understanding
integrable boundaries in statistical mechanics, conformal boundary
conditions in rational conformal field theories and the intimate
relations between them.  Indeed it appears that, for certain classes
of theories, all of the conformal boundary conditions can be realized
as the continuum scaling limit of integrable boundary conditions for
the associated integrable lattice models.  For simple rational
conformal field theories, such as $\hat{s\ell}(2)$ minimal theories, a
complete classification has been given~\cite{BePZ,BPPZ98,BPPZ00} of
the conformal boundary conditions.  These are labelled by nodes
$(r,a)$ of a tensor product graph $A\otimes G$ where the pair of
graphs $(A,G)$, with $G$ of $A$-$D$-$E$ type, coincide precisely with
the pairs in the $A$-$D$-$E$ classification of Cappelli, Itzykson and
Zuber~\cite{CIZ}.  Moreover, for the unitary minimal models the
physical content of the boundary conditions has been
ascertained~\cite{BehrendP} by studying the related integrable
boundary conditions of the associated $A$-$D$-$E$ lattice
models~\cite{Pas} with positive spectral parameter $u>0$.

It is highly desirable to carry out similar studies on boundary
conditions for other classes of conformal field theories and their
associated lattice models such as the superconformal and higher fusion
models, higher rank models and so on.  In this paper we study the
boundary conditions of ${\Bbb Z}_k$ parafermionic models~\cite{ZF85,
GepQiu, Qiu, Degio1, Degio2, DistQiu, FW98, Gep99, NY00, JKOPS00}. 
This class of rational conformal field theories admits an $A$-$D$-$E$
classification~\cite{GepQiu,DistQiu} similar to the classification of
Cappelli, Itzykson and Zuber and is associated to the same $A$-$D$-$E$
lattice models but in the regime with negative spectral parameter
$u<0$.  The first two members of the $A$ series are the Ising model
($A_3$) and the hard hexagon model~\cite{Bax80} ($A_4$) which has the
same central charge as the 3-state Potts model but admits the $W_3$
algebra as the extended chiral algebra and has 6 rather than 8
conformal boundary conditions.

The layout of the paper is as follows.  In
Section~\ref{sec:Parafermions}, we review the conformal properties of
$\hat{s\ell}(2)$ ${\Bbb Z}_k$ parafermions, their central charges,
conformal weights, characters, Kac tables and modular matrices.  We
also discuss the modular invariant partition functions on the torus
and the cylinder partition functions.  In Section~\ref{sec:Lattice},
we present the lattice realizations of these theories.  We define the
$A$-$D$-$E$ lattice models and their integrable boundary weights,
present their finite size corrections and compute their bulk and
boundary free energies.  In Section~\ref{sec:Identify} we identify,
through numerical analysis, the integrable boundary conditions from
the lattice model with the conformal boundary conditions of
parafermions.  In Section~\ref{sec:Numerics}, we discuss our numerical
techniques and present some typical numerical data on which our
results are based.  We conclude in Section~\ref{sec:Discussion} with a
short discussion.

\section{${\Bigbb Z}_k$ Parafermions} \label{sec:Parafermions}

In this section we review the conformal properties of $\hat{s\ell}(2)$
${\Bbb Z}_k$ parafermions.  In particular, we review the $A$-$D$-$E$
classifications~\cite{GepQiu,DistQiu} of torus partition functions and
present general expressions for the cylinder partition functions
following the general framework of Behrend, Pearce, Petkova and
Zuber~\cite{BPPZ00} on the assumption that the conformal boundary
conditions are labelled by nodes of the tensor product graph
\begin{equation}
G\otimes {\Bbb Z}_{2k}
\end{equation}
where the Coxeter number of $G$ is $g=k+2$ and ${\Bbb Z}_{2k}$ denotes
the {\em oriented} cyclic graph with $2k$ nodes.

\subsection{Conformal data} \label{sec:Conformal}
The $\hat{s\ell}(2)$ ${\Bbb Z}_k$ parafermionic conformal field theories
are $\hat{u}(1)$ cosets of the level $k$, $\hat{s\ell}(2)$ WZW
models
\begin{equation}
\frac{\hat{s\ell}(2)_k}{\hat{u}(1)}
\end{equation}
with central charges
\begin{equation}
c=\frac{3k}{k+2}-1=\frac{2(k-1)}{k+2}
\end{equation}
The conformal weights of primary fields $\phi_{(\ell,m)}$ are given by
\begin{equation}
\Delta_{(\ell,m)}=\frac{\ell(\ell+2)}{4(k+2)}-\frac{m^2}{4k},\qquad
0\le\ell\le k,\quad 0\le |m|\le \ell,\quad
\ell-m\in2{\Bbb Z}
\label{eq:confDims}
\end{equation}
and the associated level $k$ characters
\begin{equation}
\chi_{(\ell,m)}(q)=\eta(q)\; c^\ell_m(q)
\end{equation}
are given by the string functions~\cite{DistQiu}
\begin{eqnarray}
c^\ell_m(q)&=&\frac{q^{-\frac{c-2}{24}+\Delta_{(\ell,m)}}}{\eta \,^3(q)}
\sum_{r,s=0}^\infty (-1)^{r+s}q^{s(s+1)/2+r(r+1)/2+rs(k+1)}\non\\
&&\quad\times\left[q^{s(\ell -m)/2+r(\ell +m)/2}-q^{k+1-\ell
+s(2k+2-\ell+m)/2+r(2k+2-\ell -m)
/2}\right]
\end{eqnarray}
where $\eta(q)=q^{\frac{1}{24}}\prod_{n=1}^{\infty}(1-q^{n})$ is the
Dedekind eta function and $q=e^{2\pi i\tau}$ is the modular
parameter.

The conformal weights and characters are extended to the grid
\begin{equation}
0\le\ell\le k,\quad 0\le m\le 2k-1,\quad
\ell-m\in2{\Bbb Z}
\end{equation}
by identifying the underlying operators
\begin{equation}		\label{eq:sym}
\phi_{(\ell,m)}=\phi_{(\ell,-m)}^*=\phi_{(\ell,m+2k)}=\phi_{(k-\ell,k-m)}^*
\end{equation}
where $*$ denotes conjugation.  The conformal weights
$\Delta_{(\ell,m)}$ and characters $\chi_{(\ell,m)}(q)$ are of course
not effected by conjugation.  The parafermionic ``Kac tables" for the
critical Ising ($G=A_3$, $k=2$), critical 3-state Potts ($G=A_4$,
$k=3$) and ($G=A_5$, $k=4$) models are shown in
Table~\ref{tbl:ConfGrids}.  Each operator appears twice in these
grids.  To remove this redundancy we could fix a fundamental domain to
be the left half of the Kac table.  But in order to stress the
underlying structure of the fusion algebras we will sometimes make
other choices, such as the even sub-lattice with $\ell$ and $m$ even.

\begin{table}[p]
\nc{\spos}[2]{\makebox(0,0)[#1]{${#2}$}}
\nc{\sm}[1]{{\scriptstyle #1}}
\setlength{\unitlength}{9mm}
\begin{center}
\vspace{0.35in}
\begin{picture}(4,3)
\put(2.0,4.0){\spos{}{A_3\quad (k=2)}}
\multiput(0,0)(1,0){5}{\line(0,1){3}}
\multiput(0,0)(0,1){4}{\line(1,0){4}}
\put(2.02,0){\line(0,1){3}}
\put(-1,.5){\line(0,1){2}}
\multiput(-1,.5)(0,1){3}{\spos{}{\bullet}}
\put(-.3,-1){\line(1,0){4.6}}
\multiput(.5,-1)(1,0){4}{\spos{}{\bullet}}
\multiput(-.05,-1)(1,0){5}{\spos{}{>}}
\put(0.5,0.5){\spos{}{0}}
\put(2.5,0.5){\spos{}{\frac{1}{2}}}
\put(1.5,1.5){\spos{}{\frac{1}{16}}}
\put(3.5,1.5){\spos{}{\frac{1}{16}}}
\put(0.5,2.5){\spos{}{\frac{1}{2}}}
\put(2.5,2.5){\spos{}{0}}
\put(0.5,-0.5){\spos{}{0}}
\put(1.5,-0.5){\spos{}{1}}
\put(2.5,-0.5){\spos{}{2}}
\put(3.5,-0.5){\spos{}{3}}
\put(5,-0.5){\spos{}{m\in \Bbb Z_4}}
\put(-0.5,0.5){\spos{}{0}}
\put(-0.5,1.5){\spos{}{1}}
\put(-0.5,2.5){\spos{}{2}}
\put(-0.5,3.5){\spos{}{\ell\in A_{3}}}
\end{picture}
\vspace{0.35in}
\end{center}
\begin{center}
\vspace{0.55in}
\begin{picture}(6,4)
\put(3.0,5.0){\spos{}{A_4\quad (k=3)}}
\multiput(0,0)(1,0){7}{\line(0,1){4}}
\multiput(0,0)(0,1){5}{\line(1,0){6}}
\put(3.02,0){\line(0,1){4}}
\put(2,2){\spos{}{*}}
\put(-1,.5){\line(0,1){3}}
\multiput(-1,.5)(0,1){4}{\spos{}{\bullet}}
\put(-.3,-1){\line(1,0){6.6}}
\multiput(.5,-1)(1,0){6}{\spos{}{\bullet}}
\multiput(-.05,-1)(1,0){7}{\spos{}{>}}
\put(0.5,0.5){\spos{}{0}}
\put(2.5,0.5){\spos{}{\frac{2}{3}^*}}
\put(4.5,0.5){\spos{}{\frac{2}{3}}}
\put(1.5,1.5){\spos{}{\frac{1}{15}}}
\put(3.5,1.5){\spos{}{\frac{2}{5}}}
\put(5.5,1.5){\spos{}{\frac{1}{15}^*}}
\put(0.5,2.5){\spos{}{\frac{2}{5}}}
\put(2.5,2.5){\spos{}{\frac{1}{15}^*}}
\put(4.5,2.5){\spos{}{\frac{1}{15}}}
\put(1.5,3.5){\spos{}{\frac{2}{3}}}
\put(3.5,3.5){\spos{}{0}}
\put(5.5,3.5){\spos{}{\frac{2}{3}^*}}
\put(0.5,-0.5){\spos{}{0}}
\put(1.5,-0.5){\spos{}{1}}
\put(2.5,-0.5){\spos{}{2}}
\put(3.5,-0.5){\spos{}{3}}
\put(4.5,-0.5){\spos{}{4}}
\put(5.5,-0.5){\spos{}{5}}
\put(7,-0.5){\spos{}{m\in \Bbb Z_{6}}}
\put(-0.5,0.5){\spos{}{0}}
\put(-0.5,1.5){\spos{}{1}}
\put(-0.5,2.5){\spos{}{2}}
\put(-0.5,3.5){\spos{}{3}}
\put(-0.5,4.5){\spos{}{\ell\in A_{4}}}
\end{picture}
\vspace{0.35in}
\end{center}
\begin{center}
\vspace{0.55in}
\begin{picture}(8,5)
\put(4.0,6.0){\spos{}{A_5\quad (k=4)}}
\multiput(0,0)(1,0){9}{\line(0,1){5}}
\multiput(0,0)(0,1){6}{\line(1,0){8}}
\put(4.02,0){\line(0,1){5}}
\put(2.5,2.5){\spos{}{*}}
\put(-1,.5){\line(0,1){4}}
\multiput(-1,.5)(0,1){5}{\spos{}{\bullet}}
\put(-.3,-1){\line(1,0){8.6}}
\multiput(.5,-1)(1,0){8}{\spos{}{\bullet}}
\multiput(-.05,-1)(1,0){9}{\spos{}{>}}
\put(0.5,0.5){\spos{}{0}}
\put(2.5,0.5){\spos{}{\frac{3}{4}}}
\put(4.5,0.5){\spos{}{1}}
\put(6.5,0.5){\spos{}{\frac{3}{4}^{*}}}
\put(1.5,1.5){\spos{}{\frac{1}{16}}}
\put(3.5,1.5){\spos{}{\frac{9}{16}^{*}}}
\put(5.5,1.5){\spos{}{\frac{9}{16}}}
\put(7.5,1.5){\spos{}{\frac{1}{16}^{*}}}
\put(0.5,2.5){\spos{}{\frac{1}{3}}}
\put(2.5,2.5){\spos{}{\frac{1}{12}}}
\put(4.5,2.5){\spos{}{\frac{1}{3}}}
\put(6.5,2.5){\spos{}{\frac{1}{12}}}
\put(1.5,3.5){\spos{}{\frac{9}{16}}}
\put(3.5,3.5){\spos{}{\frac{1}{16}^{*}}}
\put(5.5,3.5){\spos{}{\frac{1}{16}}}
\put(7.5,3.5){\spos{}{\frac{9}{16}^{*}}}
\put(0.5,4.5){\spos{}{1}}
\put(2.5,4.5){\spos{}{\frac{3}{4}^{*}}}
\put(4.5,4.5){\spos{}{0}}
\put(6.5,4.5){\spos{}{\frac{3}{4}}}
\put(0.5,-0.5){\spos{}{0}}
\put(1.5,-0.5){\spos{}{1}}
\put(2.5,-0.5){\spos{}{2}}
\put(3.5,-0.5){\spos{}{3}}
\put(4.5,-0.5){\spos{}{4}}
\put(5.5,-0.5){\spos{}{5}}
\put(6.5,-0.5){\spos{}{6}}
\put(7.5,-0.5){\spos{}{7}}
\put(9,-0.5){\spos{}{m\in \Bbb Z_{8}}}
\put(-0.5,0.5){\spos{}{0}}
\put(-0.5,1.5){\spos{}{1}}
\put(-0.5,2.5){\spos{}{2}}
\put(-0.5,3.5){\spos{}{3}}
\put(-0.5,4.5){\spos{}{4}}
\put(-0.5,5.5){\spos{}{\ell\in A_{5}}}
\end{picture}
\vspace{1cm} \end{center}\caption{Conformal grids of conformal weights
for critical Ising $(A_3,{\smBbb Z}_4)$, $W_3$ $(A_4,{\smBbb Z}_6)$
and critical $(A_{5},{\smBbb Z}_{8})$ models.  The conjugation
$(l,m)\mapsto (k-l,k-m)$ is realized as a central inversion symmetry
through the point shown as $*$.}
\label{tbl:ConfGrids}
\end{table}

Under modular transformations, the level $k$ parafermionic characters
transform as
\begin{equation}
\chi_{(\ell,m)}(e^{2\pi i\tau})
=\sum_{(\ell'.m')}S_{(\ell,m)}{}
^{(\ell',m')}\;\chi_{(\ell',m')}(e^{-2\pi i/\tau})
\end{equation}
where the modular matrix $S$ satisfies
\begin{equation}
S^\dagger=S^{-1}, \quad S^4=I,\quad S^2=C
\end{equation}
and $C$ is the matrix implementing conjugation.  If $(\ell,m)$ ranges
over the full conformal grid then $S=S_{WZW}\otimes S_{\mbox{$\Bbb
Z$}_{2k}}$ is just the tensor product of the $\hat{s\ell}(2)_k$
Wess-Zumino-Witten modular matrix and the $\hat{s\ell}(2k)_1$ modular
matrix.  Restricting to a fundamental domain, the entries of the
$k(k+1)/2\times k(k+1)/2$ modular matrix are simply renormalized by a
factor of 2 and given explicitly by
\begin{equation}
S_{(\ell,m)}{}^{(\ell',m')}=
\frac1{\sqrt{k(k+2)}}\sin\frac{\pi(\ell+1)(\ell'+1)}{(k+2)}\;\exp\frac
{\pi imm'}{k}
\end{equation}

\goodbreak
\subsection{${\BBbb Z}_k$ parafermions on a torus}
\label{sec:ParaTorus}
The $\hat{s\ell}(2)$ ${\Bbb Z}_k$ parafermions on a torus have been
classified by Gepner and Qiu~\cite{GepQiu} using modular invariance. 
They distinguish principal and non-principal cases and their results
are summarised in Tables~\ref{PrincClass} and \ref{NonPrincClass}. 
Here we will be primarily concerned with the principal cases for which
the identification with lattice models is clear.  We conjecture that
non-principal models are associated with fused models~\cite{KP92,
BPO'B96}.

\begin{table}[p]
$$
\begin{array}{ll}
(A_{k+1},{\Bbb Z}_{2k}):& \frac12
\disp{\sum_{m=0}^{2k-1}\sum_{\ell=0}^k |\chi_{(\ell,m)}|^2}\\ &\\
(D_{2n+2},{\Bbb Z}_{8n}):
& \disp{
\sum_{m=0}^{2k-1}
\left\{
\sum_{\topped{\ell=0}{\ell\mbox{\tiny   even}}}^{k/2-1}
\left\{|\chi_{(\ell,m)}|^2+
\chi_{(k-\ell,m)}{\;\overline{\chi}_{(\ell,m)}}\right\}
+2|\chi_{(k/2,m)}|^2
\right\}
} \\
(k=4n)&=\frac12
\disp{\sum_{m=0}^{2k-1}
\sum_{\topped{\ell=0}{\ell
\mbox{\tiny  even}}}^{k}
       |\chi_{(\ell,m)}+\chi_{(k-\ell,m)}|^2}\\ &\\
(D_{2n+1},{\Bbb Z}_{8n-4}): &\frac12
\disp{\sum_{m=0}^{2k-1}\left\{
\sum_{\topped{\ell=0}{\ell \mbox{\tiny\  even}}}^{k}
|\chi_{(\ell,m)}|^2+\sum_{\topped{\ell=0}{\ell
\mbox{\tiny\  odd}}}^{k} \chi_{(k-\ell,m)}{\;\overline
\chi_{(\ell,m)}}\right\}}
\\ (k=4n-2)&\\
&\\
(E_6,{\Bbb Z}_{20}):&\frac12
\disp{\sum_{m=0}^{2k-1}
\left\{|\chi_{(0,m)}+\chi_{(6,m)}|^2+|\chi_{(3,m)}+\chi_{(7,m)}|^2+
|\chi_{(4,m)}+\chi_{(10,m)}|^2\right\}}\\
(k=10)&\\ &\\
(E_7,{\Bbb Z}_{32}):&\frac12
\disp{\sum_{m=0}^{2k-1}
\Bigl\{|\chi_{(0,m)}+\chi_{(16,m)}|^2+|\chi_{(4,m)}+\chi_{(12,m)}|^2+
|\chi_{(6,m)}+\chi_{(10,m)}|^2}\\
(k=16)&\disp{\qquad\qquad+|\chi_{(8,m)}|^2
+(\chi_{(2,m)}+\chi_{(14,m)})\overline \chi_{(8,m)}+\chi_{(8,m)}(\overline
\chi_{(2,m)}+\overline \chi_{(14,m)})\Bigr\}}\\ \\
(E_8,{\Bbb Z}_{56}):&\frac12
\disp{\sum_{m=0}^{2k-1}\Bigl\{
|\chi_{(0,m)}+\chi_{(10,m)}+\chi_{(18,m)}+\chi_{(28,m)}|^2}\\
(k=28)&\disp{\qquad\qquad+|\chi_{(6,m)}+\chi_{(12,m)}
+\chi_{(16,m)}+\chi_{(22,m)}|^2\Bigr\}}
\end{array}
$$
\caption{The $A$-$D$-$E$ classification of modular invariant
partition functions for
principal ${\Bbb Z}_k$ theories.}\label{PrincClass}
\end{table}

\begin{table}[p]
$$
\begin{array}{ll}
       (A_{k+1},{\Bbb Z}_{2\beta}):& \frac14
\disp{\sum_{x\in{\Bbb Z}_{2\beta}, y\in{\Bbb
Z}_{2\alpha}}\sum_{\ell=0}^k \chi_{(\ell,{\alpha^+})} \overline
\chi_{(\ell,{\alpha^-})}}\\ &\\
(D_{2n+2},{\Bbb Z}_{2\beta}):
&\frac18
\disp{\sum_{x\in{\Bbb Z}_{2\beta}, y\in{\Bbb Z}_{2\alpha}}
\sum_{\topped{\ell=0}{\ell
\mbox{
\tiny\   even}}}^{k}(\chi_{(\ell,\alpha^+)}+
\chi_{(k-\ell,\alpha^+)}) (\overline
\chi_{(\ell,\alpha^-)}+ \overline \chi_{(k-\ell,\alpha^-)})} \\ (k=4n)&\\ &\\
(D_{2n+1},{\Bbb Z}_{2\beta}): &\frac14
\disp{\sum_{x\in{\Bbb
Z}_{2\beta}, y\in{\Bbb Z}_{2\alpha}}\left\{
\sum_{\topped{\ell=0}{\ell \mbox{\tiny\  even}}}^{k}  \chi_{(\ell,\alpha^+)}
\overline \chi_{(\ell,\alpha^-)} +\sum_{\topped{\ell=0}{\ell
\mbox{\tiny\  odd}}}^{k} \chi_{(k-\ell,\alpha^+)}{\;\overline
\chi_{(\ell,\alpha^-)}}\right\}}
\\ (k=4n-2)&\\  \\
(E_6,{\Bbb Z}_{2\beta}):&\frac14
     \disp{\sum_{x\in{\Bbb Z}_{2\beta},
y\in{\Bbb Z}_{2\alpha}}
\Bigl\{(\chi_{(0,\alpha^+)}+\chi_{(6,\alpha^+)})(\overline
\chi_{(0,\alpha^-)}+\overline
\chi_{(6,\alpha^-)})}\\
(k=10)&{}+\disp{(\chi_{(3,\alpha^+)}+\chi_{(7,\alpha^+)})(\overline
\chi_{(3,\alpha^-)}+\overline
\chi_{(7,\alpha^-)})+(\chi_{(4,\alpha^+)}+\chi_{(10,\alpha^+)})(\overline
\chi_{(4,\alpha^-)}+\overline \chi_{(10,\alpha^-)})\Bigr\}}\\  &\\
(E_7,{\Bbb Z}_{2\beta}):&\frac14 \!\!
\disp{\sum_{x\in{\Bbb Z}_{2\beta}, y\in{\Bbb Z}_{2\alpha}}\!\!\!\!
\Bigl\{(\chi_{(0,\alpha^+)}+\chi_{(16,\alpha^+)})(\overline
\chi_{(0,\alpha^-)}+\overline \chi_{(16,\alpha^-)})}\\
&{}+\disp{(\chi_{(4,\alpha^+)}+\chi_{(12,\alpha^+)})(\overline
\chi_{(4,\alpha^-)}+\overline \chi_{(12,\alpha^-)})+
(\chi_{(6,\alpha^+)}+\chi_{(10,\alpha^+)})(\overline
\chi_{(6,\alpha^-)}+\overline
\chi_{(10,\alpha^-)})}\\
       (k=16)&\disp{\qquad\qquad+\chi_{(8,\alpha^+)}\overline 
\chi_{(8,\alpha^-)}
+(\chi_{(2,\alpha^+)}+\chi_{(14,\alpha^+)})\overline
\chi_{(8,\alpha^-)}+\chi_{(8,\alpha^+)}(\overline \chi_{(2,\alpha^-)}+\overline
\chi_{(14,\alpha^-)})\Bigr\}}\\& \\
       (E_8,{\Bbb Z}_{2\beta}):&\frac14 \!\!
\disp{\sum_{x\in{\Bbb Z}_{2\beta}, y\in{\Bbb Z}_{2\alpha}}\!\!\!\!\Bigl\{
(\chi_{(0,\alpha^+)}\!+\!\chi_{(10,\alpha^+)}\!+\!\chi_{(18,\alpha^+)}
\!+\!\chi_{(28,\alpha^+)})(\overline
\chi_{(0,\alpha^-)}\!+\!\overline \chi_{(10,\alpha^-)}\!+\!\overline
\chi_{(18,\alpha^-)}\!+\!\overline \chi_{(28,\alpha^-)})}\\
(k=28)&\disp{\qquad\qquad+(\chi_{(6,\alpha^+)}\!+\!\chi_{(12,\alpha^+)}
\!+\!\chi_{(16,\alpha^+)}\!+\!\chi_{(22,\alpha^+)})
(\overline \chi_{(6,\alpha^-)}\!+\!\overline \chi_{(12,\alpha^-)}\!+\!\overline
\chi_{(16,\alpha^-)}\!+\!\overline \chi_{(22,\alpha^-)})\Bigr\}}
\\ &\end{array}
$$
\caption{The $A$-$D$-$E$ classification of modular invariant
partition functions for
non-principal ${\Bbb Z}_k$ theories. Here $\alpha\ne 1$ is a divisor
of $k=\alpha\beta$ and $\alpha^+=\alpha x +\beta y$,
$\alpha^-=\alpha x-\beta y$.}\label{NonPrincClass}
\end{table}

\goodbreak
\subsection{${\BBbb Z}_{k}$ parafermions on a cylinder}
\label{sec:ParaCyl}
In this section we consider the ${\Bbb Z}_{k}$ parafermion models on a
cylinder. We restrict our
attention to the principal cases $(G,{\Bbb Z}_{2k})$ where $G$ is
of $A$, $D$ or $E$ type.
When $G$ is of $A$ type, the theory is diagonal and the cylinder
partition functions are
\begin{equation}
Z_{(\ell_1,m_1)|(\ell_2,m_2)}(q)=\sum_{(\ell,m)} N_{(\ell,m)\;
(\ell_1,m_1)}{}^{(\ell_2,m_2)} \chi_{(\ell,m)}(q)\label{Verlinde}
\end{equation}
where $(\ell_1,m_1)$ labels the boundary condition on the left,
$(\ell_2,m_2)$ labels the boundary condition on the right and the
fusion coefficients are given by the Verlinde
formula~\cite{Ve}
\begin{equation}
N_{(\ell,m)\; (\ell_1,m_1)}{}^{(\ell_2,m_2)}= \sum_{(\ell',m')}
\frac{S_{(\ell,m)}{}^{(\ell',m')}\; S_{(\ell_1,m_1)}{}^{(\ell',m')}\;
{S_{(\ell_2,m_2)}{}^{(\ell',m')}\,}^*}{S_{(0,0)}{}^{(\ell',m')}}\in{\Bbb
N}
\end{equation}
The parafermionic cylinder partition functions for $(A_{4},{\Bbb
Z}_{6})$ are shown in Table~\ref{tbl:cylinderA4}.  The characters
associated to conjugate primary fields, although equal, have not been
used so that this table encodes the complete fusion algebra with the
conjugate operators shown in the first column.

\begin{table}[htb]
$$\begin{array}{cc}&\text{Right}\\[.5cm] \text{Left}&\begin{array}{c|cccccc}
(\ell,m)&(0,0)&(2,0)&(0,2)&(2,2)&(0,4)&(2,4)\\
\hline
(0,0)&
          \, \chi_{(0,0)} &
          \, \chi_{(0,2)}&
          \, \chi_{(0,4)}&
          \, \chi_{(2,0)} &
          \, \chi_{(2,2)} &
          \, \chi_{(2,4)} \\
(0,2)&
          \, \chi_{(0,4)}  &
          \, \chi_{(0,0)} &
          \, \chi_{(0,2)}   &
          \,  \chi_{(2,4)}   &
          \, \chi_{(2,0)}   &
          \,  \chi_{(2,2)}\\
(0,4)&
          \, \chi_{(0,2)}   &
          \, \chi_{(0,4)}   &
          \, \chi_{(0,0)}   &
          \, \chi_{(2,2)}  &
          \, \chi_{(2,4)}   &
          \, \chi_{(2,0)}   \\
(2,0)&
          \, \chi_{(2,0)}   &
          \, \chi_{(2,2)} &
          \, \chi_{(2,4)}   &
          \, \chi_{(0,0)} +\chi_{(2,0)} \, &
          \, \chi_{(0,2)} +\chi_{(2,2)} \, &
          \, \chi_{(0,4)} +\chi_{(2,4)} \, \\
(2,2)&
          \, \chi_{(2,4)} &
          \, \chi_{(2,0)}   &
          \, \chi_{(2,2)}   &
          \, \chi_{(0,4)} +\chi_{(2,4)} \, &
          \, \chi_{(0,0)} +\chi_{(2,0)} \, &
          \, \chi_{(0,2)} +\chi_{(2,2)} \, \\
(2,4)&
          \, \chi_{(2,2)}   &
          \, \chi_{(2,4)},&
          \, \chi_{(2,0)}   &
          \, \chi_{(0,2)} +\chi_{(2,2)} \, &
          \, \chi_{(0,4)} +\chi_{(2,4)} \, &
          \, \chi_{(0,0)} +\chi_{(2,0)} \,
\end{array}\end{array}$$
\caption{ Cylinder partition functions $Z_{L|R}$ for the $W_3$ model
$(A_4,\Bbb Z_6)\sim (T_2,\Bbb Z_3)$.
}
     					\label{tbl:cylinderA4}
\end{table}

If $G$ is of $D$ or $E$ type, the boundary conditions are labelled by
$(a,m)\in(G,{\Bbb Z}_{2k})$ and the cylinder partition functions are
given by~\cite{BPPZ00}:
\begin{equation}
Z_{(a_1,m_1)|(a_2,m_2)}(q)=\sum_{(\ell,m)}
V_{(\ell,m)\; (a_1,m_1)}{}^{(a_2,m_2)} \chi_{(\ell,m)}(q)
\end{equation}
where the sum is restricted to a fundamental domain of the Kac
table. The intertwiners are
\begin{equation}
V_{(\ell,m)\; (a_1,m_1)}{}^{(a_2,m_2)}=
\sum_{(a',m')} \frac{S_{(\ell,m)}{}^{(a'-1,m')}\;
\Psi_{(a_1,m_1)}{}^{(a',m')}\;
{\Psi_{(a_2,m_2)}{}^{(a',m')}}^*}{S_{(0,0)}{}^{(a'-1,m')}}\in{\Bbb N}
\label{eq:Verlinde}
\end{equation}
where the sum is over $a'\in \mbox{Exp}(G)$ the Coxeter exponents of
$G$ and $m'\in{\Bbb Z}_{2k}$.  The eigenvectors of the adjacency
matrix of the graph $G\otimes{\Bbb Z}_{2k}$ are given by the tensor
product
\begin{equation}
\Psi_{(a,m)}{}^{(a',m')}=\psi_a{}^{a'}\; \exp{\frac{\pi imm'}{k}}
\end{equation}
where $\psi^{a'}$ is the eigenvector, labelled by $a'\in
\mbox{Exp}(G)$, of the adjacency matrix $G$.

The intertwiner can alternatively be defined as the tensor product of
intertwiners for each graph
\begin{equation}
        V_{(\ell,m)\; (a_1,m_1)}{}^{(a_2,m_2)}=
        V^{{}^G}_{\ell+1,\;a_{1}}{}^{a_{2}}\;\;
        V^{{}^{Z}}_{m+1,\;m_{1}}{}^{m_{2}}
        \label{eq:intertwinProd}
\end{equation}
where the intertwiners for $G$ and $Z={\Bbb Z}_{2k}$ are defined by a
formulas similar to (\ref{eq:Verlinde}).  The intertwiner for the
graph $G$ satisfies the recurrence relation
\begin{equation}
V^{{}^G}_{i}:=V^{{}^G}_{2}\cdot V^{{}^G}_{i-1}-V^{{}^G}_{i-2}
\end{equation}
where $V^{{}^G}_{1}=I$ and $V^{{}^G}_{2}=G$, the adjacency matrix of
$G$.  The intertwiner for ${\Bbb Z}_{2k}$ is simply a power of the
adjacency matrix $V^{{}^{Z}}_{m}=({\Bbb Z}_{2k})^{m}$ of the directed
graph ${\Bbb Z}_{2k}$.

The parafermionic cylinder partition functions for $(A_{5},{\Bbb
Z}_{8})$ and $(D_{4},{\Bbb Z}_{8})$ are shown in
Tables~\ref{tbl:cylinderA5}-\ref{tbl:cylinderD4}.  In the $A_{5}$ case
the characters associated with conjugate fields have been kept
separated in order to encode the complete fusion algebra, but have
been identified in the $D_{4}$ case to make the $\Bbb Z_{3}$ symmetry
apparent when these $(D_{4},{\Bbb Z}_{8})$ partition functions are
written in terms of the fundamental string functions.

In the general $D$ and $E$ cases, these partition functions are best
understood in terms of extended (block) characters~\cite{BPPZ00}. 
These extended characters $\hat\chi_{(a,m)}$ are linear forms in the
$(A_{L},\Bbb Z_{2k})$ characters involving the fundamental intertwiner
\begin{equation}
        \hat\chi_{(a,m)}(q):=\sum_{(\ell,n)}
V_{(\ell,n)\;(0,0)}{}^{(a,m)}\chi_{(\ell,n)}(q)
        \label{eq:block}
\end{equation}
where the sum is over a fundamental domain.  Note that for the
 $(D_{\frac k 2 +2}, \Bbb Z_{2k})$ models the conformal dimensions of the two primary
fields appearing in the extended characters differ by integers or half
integers
\begin{equation}
\Delta_{(\ell,m)}-\Delta_{(k-\ell,m)}
=\frac{\ell(\ell+2)-(k-\ell)(k-\ell+2)}{4(k+2)}
=\frac{\ell-\frac{k}{2}}{2}
\label{eq:confDimsDif}
\end{equation}
Note also that the parafermionic symmetries~(\ref{eq:sym}) imply that
these models are in fact $\Bbb Z_{k}$
symmetric.

\begin{landscape}
\begin{table}[tbp]
        \centering
\newcolumntype{t}{>{\scriptscriptstyle}c}
        $$\begin{array}{t|tttttttttt}
(\ell,m)&(0,0)&(0,2)&(0,4)&(0,6)&(2,0)&(2,2)&(1,1)&(1,3)&(1,5)&(1,7)\\
\hline
(0,0)&
	\, \chi_{(0,0)} &
	\, \chi_{(0,2)}&
	\, \chi_{(0,4)} &
	\, \chi_{(0,6)}&
	\, \chi_{(2,0)}&
	\, \chi_{(2,2)}&
	\, \chi_{(1,1)}&
	\, \chi_{(1,3)}&
	\, \chi_{(1,5)}&
	\, \chi_{(1,7)}  \\
(0,2)&
          \, \chi_{(0,6)}   &
          \, \chi_{(0,0)}   &
          \, \chi_{(0,2)}   &
          \, \chi_{(0,4)}   &
          \, \chi_{(2,2)}   &
          \, \chi_{(2,0)}   &
          \, \chi_{(1,7)}   &
          \, \chi_{(1,1)}   &
          \, \chi_{(1,3)}   &
          \, \chi_{(1,5)}   \\
(0,4)&
          \, \chi_{(0,4)}   &
          \, \chi_{(0,6)}   &
          \, \chi_{(0,0)}   &
          \, \chi_{(0,2)}   &
          \, \chi_{(2,0)}   &
          \, \chi_{(2,2)}   &
          \, \chi_{(1,5)}   &
          \, \chi_{(1,7)}   &
          \, \chi_{(1,1)}   &
          \, \chi_{(1,3)}    \\
(0,6)&
          \, \chi_{(0,2)}   &
          \, \chi_{(0,4)}   &
          \, \chi_{(0,6)}   &
          \, \chi_{(0,0)}   &
          \, \chi_{(2,2)}   &
          \, \chi_{(2,0)}   &
          \, \chi_{(1,3)}   &
          \, \chi_{(1,5)}   &
          \, \chi_{(1,7)}   &
          \, \chi_{(1,1)}   \\
(2,0)&
          \, \chi_{(2,0)}   &
          \, \chi_{(2,2)}   &
          \, \chi_{(2,0)}   &
          \, \chi_{(2,2)}  &
          \, \chi_{(0,0)}+ \chi_{(2,0)}   + \chi_{(4,0)}     \, &
          \, \chi_{(0,2)}+ \chi_{(2,2)}   + \chi_{(4,2)}     \, &
          \,  \chi_{(1,1)}   + \chi_{(1,5)}     \, &
          \,  \chi_{(1,3)}   + \chi_{(1,7)}     \, &
          \,  \chi_{(1,1)}   + \chi_{(1,5)}     \, &
          \,  \chi_{(1,3)}   + \chi_{(1,7)}     \, \\
(2,2)&
          \, \chi_{(2,2)}  &
          \, \chi_{(2,0)}   &
          \, \chi_{(2,2)}   &
          \, \chi_{(2,0)}   &
          \, \chi_{(0,2)}+ \chi_{(2,2)}   + \chi_{(4,2)}     \, &
          \, \chi_{(0,0)}+ \chi_{(2,0)}   + \chi_{(4,0)}     \, &
          \,  \chi_{(1,3)}   + \chi_{(1,7)}       \, &
          \,  \chi_{(1,1)}   + \chi_{(1,5)}   \, &
          \,  \chi_{(1,3)}   + \chi_{(1,7)}     \, &
          \,  \chi_{(1,1)}   + \chi_{(1,5)}     \,
\\
     (1,1)&
           \, \chi_{(1,7)}  &
           \, \chi_{(1,1)}  &
           \, \chi_{(1,3)}   &
           \, \chi_{(1,5)}   &
          \,  \chi_{(1,3)}  + \chi_{(1,7)} \,&
          \,  \chi_{(1,1)}   + \chi_{(1,5)}     \, &
          \,  \chi_{(0,0)}   + \chi_{(2,0)} \,&
          \,  \chi_{(0,2)}   + \chi_{(2,2)} \,&
          \,  \chi_{(0,4)}  + \chi_{(2,0)}     \, &
          \,  \chi_{(0,6)}  + \chi_{(2,2)}      \\
(1,3)&
          \, \chi_{(1,5)}   &
          \, \chi_{(1,7)}  &
          \, \chi_{(1,1)}  &
          \, \chi_{(1,3)}   &
           \,  \chi_{(1,1)}   + \chi_{(1,5)}     \, &
           \,  \chi_{(1,3)}  + \chi_{(1,7)} \,&
           \,  \chi_{(0,6)}  + \chi_{(2,2)} \,&
           \,  \chi_{(0,0)}   + \chi_{(2,0)} \,&
           \,  \chi_{(0,2)}   + \chi_{(2,2)} \, &
           \,  \chi_{(0,4)}  + \chi_{(2,0)}
     \\
(1,5)&
          \, \chi_{(1,3)}   &
          \, \chi_{(1,5)}   &
          \, \chi_{(1,7)}  &
          \, \chi_{(1,1)}  &
           \,  \chi_{(1,3)}  + \chi_{(1,7)} \,&
           \,  \chi_{(1,1)}   + \chi_{(1,5)}     \, &
           \,  \chi_{(0,4)}  + \chi_{(2,0)}     \, &
           \,  \chi_{(0,6)}  + \chi_{(2,2)} \,&
           \,  \chi_{(0,0)}   + \chi_{(2,0)} \,&
           \,  \chi_{(0,2)}   + \chi_{(2,2)}
\\
(1,7)&
          \, \chi_{(1,1)}  &
          \, \chi_{(1,3)}   &
          \, \chi_{(1,5)}   &
          \, \chi_{(1,7)}  &
           \,  \chi_{(1,1)}   + \chi_{(1,5)}     \, &
           \,  \chi_{(1,3)}  + \chi_{(1,7)} \,&
           \,  \chi_{(0,2)}   + \chi_{(2,2)} \,&
           \,  \chi_{(0,4)}  + \chi_{(2,0)}     \, &
           \,  \chi_{(0,6)}  + \chi_{(2,2)}  \,&
           \,  \chi_{(0,0)}   + \chi_{(2,0)}
\end{array}$$
\caption{Cylinder partition functions $Z_{L|R}$ for $(A_5,\Bbb Z_8)$. 
The rows are labelled by the left boundary condition and the columns
by the right boundary condition.  The table is in fact symmetric due
to conjugation symmetry.
     }
        \label{tbl:cylinderA5}
\end{table}
\begin{table}[tbp]
        \centering
\newcolumntype{t}{>{\scriptscriptstyle}c}
        $$\begin{array}{t|tttttt}
(a,m)&(A,0)&(A,2)&(B,0)&(B,2)&(O,1)&(O,3)\\
\hline
(A,0)&\chi_{(0,0)}+\chi_{(4,0)}& 2\chi_{(0,2)} &
\chi_{(2,0)}& \chi_{(2,2)}& \chi_{(1,1)}+\chi_{(1,3)} &
\chi_{(1,1)}+\chi_{(1,3)}\\
(A,2)&2\chi_{(0,2)}& \chi_{(0,0)}+\chi_{(4,0)} &
\chi_{(2,2)}& \chi_{(2,0)}& \chi_{(1,1)}+\chi_{(1,3)}&
\chi_{(1,1)}+\chi_{(1,3)} \\
(B,0)&\chi_{(2,0)}& \chi_{(2,2)} & \chi_{(0,0)}+\chi_{(4,0)}& 2\chi_{(0,2)} &
\chi_{(1,1)}+\chi_{(1,3)} & \chi_{(1,1)}+\chi_{(1,3)}\\
(B,2) & \chi_{(2,2)}& \chi_{(2,0)}& 2\chi_{(0,2)}&\chi_{(0,0)}+\chi_{(4,0)}&
\chi_{(1,1)}+\chi_{(1,3)} & \chi_{(1,1)}+\chi_{(1,3)}\\
(O,1)& \chi_{(1,1)}+\chi_{(1,3)} & \chi_{(1,1)}+\chi_{(1,3)}&
\chi_{(1,1)}+\chi_{(1,3)} & \chi_{(1,1)}+\chi_{(1,3)}&
\chi_{(0,0)}+2\chi_{(2,0)}+ \chi_{(4,0)}&
2\chi_{(0,2)}+2\chi_{(2,2)}\\
(O,3)& \chi_{(1,1)}+\chi_{(1,3)}& \chi_{(1,1)}+\chi_{(1,3)} &
\chi_{(1,1)}+\chi_{(1,3)}& \chi_{(1,1)}+\chi_{(1,3)} &
2\chi_{(0,2)}+ 2\chi_{(2,2)}&
\chi_{(0,0)}+2\chi_{(2,0)}+ \chi_{(4,0)}\\
\end{array}$$
\caption{Cylinder partition functions $Z_{L|R}$ for $(D_{4},\Bbb
Z_8)$.  $O$ stands for the center point of $D_{4}$ and $A$, $B$ for
any two of the three other points.  The table is ${\Bbb Z}_3$
symmetric.
}
        \label{tbl:cylinderD4}
\end{table}

\begin{table}[p]
\nc{\spos}[2]{\makebox(0,0)[#1]{${#2}$}}
\nc{\sm}[1]{{\scriptstyle #1}}
\setlength{\unitlength}{9mm}
\begin{center}
\begin{picture}(13,5)
\put(5.0,6.0){\spos{}{A_5\quad (k=4)}}
\multiput(0,0)(1,0){9}{\line(0,1){5}}
\multiput(0,0)(0,1){6}{\line(1,0){8}}
\put(4.02,0){\line(0,1){5}}
\put(-1,.5){\line(0,1){4}}
\multiput(-1,.5)(0,1){5}{\spos{}{\bullet}}
\put(-.3,-1){\line(1,0){8.6}}
\multiput(.5,-1)(1,0){8}{\spos{}{\bullet}}
\multiput(-.05,-1)(1,0){9}{\spos{}{>}}
\put(0.5,0.5){\spos{}{0}}
\put(2.5,0.5){\spos{}{\frac{3}{4}}}
\put(4.5,0.5){\spos{}{1}}
\put(6.5,0.5){\spos{}{\frac{3}{4}^{*}}}
\put(1.5,1.5){\spos{}{\frac{1}{16}}}
\put(3.5,1.5){\spos{}{\frac{9}{16}^{*}}}
\put(5.5,1.5){\spos{}{\frac{9}{16}}}
\put(7.5,1.5){\spos{}{\frac{1}{16}^{*}}}
\put(0.5,2.5){\spos{}{\frac{1}{3}}}
\put(2.5,2.5){\spos{}{\frac{1}{12}}}
\put(4.5,2.5){\spos{}{\frac{1}{3}}}
\put(6.5,2.5){\spos{}{\frac{1}{12}}}
\put(1.5,3.5){\spos{}{\frac{9}{16}}}
\put(3.5,3.5){\spos{}{\frac{1}{16}^{*}}}
\put(5.5,3.5){\spos{}{\frac{1}{16}}}
\put(7.5,3.5){\spos{}{\frac{9}{16}^{*}}}
\put(0.5,4.5){\spos{}{1}}
\put(2.5,4.5){\spos{}{\frac{3}{4}^{*}}}
\put(4.5,4.5){\spos{}{0}}
\put(6.5,4.5){\spos{}{\frac{3}{4}}}
\put(0.5,-0.5){\spos{}{0}}
\put(1.5,-0.5){\spos{}{1}}
\put(2.5,-0.5){\spos{}{2}}
\put(3.5,-0.5){\spos{}{3}}
\put(4.5,-0.5){\spos{}{4}}
\put(5.5,-0.5){\spos{}{5}}
\put(6.5,-0.5){\spos{}{6}}
\put(7.5,-0.5){\spos{}{7}}
\put(9,-0.5){\spos{}{m\in \Bbb Z_{8}}}
\put(-0.5,0.5){\spos{}{0}}
\put(-0.5,1.5){\spos{}{1}}
\put(-0.5,2.5){\spos{}{2}}
\put(-0.5,3.5){\spos{}{3}}
\put(-0.5,4.5){\spos{}{4}}
\put(-0.5,5.5){\spos{}{\ell\in A_{5}}}
\end{picture}
\begin{picture}(5,5)
\put(2.0,6.0){\spos{}{D_4\quad (k=4)}}
\multiput(0,0)(1,0){6}{\line(0,1){3}}
\multiput(0,0)(0,1){4}{\line(1,0){5.3}}
\put(4.02,0){\line(0,1){3}}
\put(-1.5,0.5){\line(0,1){1}}
\multiput(-1.5,0.5)(0,1){2}{\spos{}{\bullet}}
\put(-0.5,2.5){\spos{}{\bullet}}
\put(-2.5,2.5){\spos{}{\bullet}}
\put(-1.5,1.5){\line(1,1){1}}
\put(-1.5,1.5){\line(-1,1){1}}
\put(-.3,-1.25){\line(1,0){5.6}}
\multiput(.5,-1.25)(1,0){5}{\spos{}{\bullet}}
\multiput(-.05,-1.25)(1,0){6}{\spos{}{>}}
\put(0.5,0.5){\spos{}{\sm{0+1}}}
\put(2.5,0.5){\spos{}{\sm{\frac{3}{4}+{\frac{3}{4}}^{*}}}}
\put(4.5,0.5){\spos{}{\sm{0+1}}}
\put(1.5,1.5){\spos{}{\sm{\frac{1}{16}+{\frac{9}{16}}}}}
\put(3.5,1.5){\spos{}{\sm{\frac{\smash{1^{*}}}{16}+\frac{\smash{9^{*}}}{16}}}}
\put(0.5,2.5){\spos{}{\sm{\frac{1}{3},\frac{1}{3}^{*}}}}
\put(2.5,2.5){\spos{}{\sm{\frac{1}{12},\frac{1}{12}^{*}}}}
\put(4.5,2.5){\spos{}{\sm{\frac{1}{3},\frac{1}{3}^{*}}}}
\put(0.5,-0.5){\spos{}{0}}
\put(1.5,-0.5){\spos{}{1}}
\put(2.5,-0.5){\spos{}{2}}
\put(3.5,-0.5){\spos{}{3}}
\put(4.5,-0.5){\spos{}{4}}
\put(6,-0.5){\spos{}{m\in \Bbb Z_{8}}}
\put(-1,0.5){\spos{}{A}}
\put(-1,1.5){\spos{}{O}}
\put(-0.5,3){\spos{}{C}}
\put(-2.5,3){\spos{}{B}}
\put(-1.5,4){\spos{}{a\in D_{4}}}
\end{picture}
\vspace{1cm}
\end{center}
\caption{Conformal grid for the $(D_{4},{\smBbb Z}_{8})$ model
resulting from the folding of the $(A_{5},{\smBbb Z}_{8})$ model.
$A,B,C$ are even, $O$ is odd.}
\label{tbl:ConfGridD4}
\end{table}

\begin{table}[tbp]
        \centering
      $$  \begin{array}{rclcrcl}
            \hat\chi_{(A,0)}&=&\chi_{(0,0)}+\chi_{(4,0)}  &\qquad\qquad&
            \hat\chi_{(A,2)}&=&\chi_{(0,2)}+\chi_{(4,2)}  \\
            \hat\chi_{(O,1)}&=&\chi_{(1,1)}+\chi_{(1,5)}  &&
            \hat\chi_{(O,3)}&=&\chi_{(1,7)}+\chi_{(1,3)}  \\
            \hat\chi_{(B,0)}&=&\chi_{(2,0)} &&
            \hat\chi_{(B,2)}&=&\chi_{(2,2)} \\
            \hat\chi_{(C,0)}&=&\chi_{(2,0)} &&
            \hat\chi_{(C,2)}&=&\chi_{(2,2)}
      \end{array}$$
\newcolumntype{t}{>{}c}
        $$\begin{array}{t|tttttttt}
(a,m)&(A,0)&(A,2)&(B,0)&(B,2)&(C,0)&(C,2)&(O,1)&(O,3)\\
\hline
(A,0)&
     \hat\chi_{(A,0)}& \hat\chi_{(A,2)} &
     \hat\chi_{(B,0)}& \hat\chi_{(B,2)}&
     \hat\chi_{(C,0)}& \hat\chi_{(C,2)}&  \hat\chi_{(O,1)}&
     \hat\chi_{(O,3)}\\
(A,2)&
     \hat\chi_{(A,2)}& \hat\chi_{(A,0)} &
     \hat\chi_{(B,2)}& \hat\chi_{(B,0)}&
     \hat\chi_{(C,2)}& \hat\chi_{(C,0)}&  \hat\chi_{(O,3)}&
     \hat\chi_{(O,1)}\\
(B,0)&
     \hat\chi_{(C,0)}&
     \hat\chi_{(C,2)}&
     \hat\chi_{(A,0)}& \hat\chi_{(A,2)}&\hat\chi_{(B,0)}&
\hat\chi_{(B,2)}&  \hat\chi_{(O,1)}&
     \hat\chi_{(O,3)}\\
(B,2) &
     \hat\chi_{(C,2)}& \hat\chi_{(C,0)}&
     \hat\chi_{(A,2)}& \hat\chi_{(A,0)}&\hat\chi_{(B,2)}&
\hat\chi_{(B,0)}&  \hat\chi_{(O,3)}&
     \hat\chi_{(O,1)}\\
(C,0)&
     \hat\chi_{(B,0)}& \hat\chi_{(B,2)} & \hat\chi_{(C,0)}&
     \hat\chi_{(C,2)}&
     \hat\chi_{(A,0)}& \hat\chi_{(A,2)}&  \hat\chi_{(O,1)}&
     \hat\chi_{(O,3)}\\
(C,2) &
     \hat\chi_{(B,2)}& \hat\chi_{(B,0)} &
     \hat\chi_{(C,2)}& \hat\chi_{(C,0)}&
     \hat\chi_{(A,2)}& \hat\chi_{(A,0)}&  \hat\chi_{(O,3)}&
     \hat\chi_{(O,1)}\\
(O,1)&
     \hat\chi_{(O,3)}& \hat\chi_{(O,1)} & \hat\chi_{(O,3)}&
     \hat\chi_{(O,1)}& \hat\chi_{(O,3)}& \hat\chi_{(O,1)}&
     \hat\chi_{(A,0)}+ \hat\chi_{(B,0)}+ \hat\chi_{(C,0)}&
     \hat\chi_{(A,2)}+ \hat\chi_{(B,2)}+ \hat\chi_{(C,2)}\\
(O,3)&
     \hat\chi_{(O,1)}& \hat\chi_{(O,3)}& \hat\chi_{(O,1)}&
     \hat\chi_{(O,3)}& \hat\chi_{(O,1)}& \hat\chi_{(O,3)} &
     \hat\chi_{(A,2)}+ \hat\chi_{(B,2)}+ \hat\chi_{(C,2)}&
     \hat\chi_{(A,0)}+ \hat\chi_{(B,0)}+ \hat\chi_{(C,0)}\\
\end{array}$$
\caption{Cylinder partition functions (fusion algebra) for
$(D_{4},\Bbb Z_8)$ in terms of extended characters.  $A$ is the
distinguished node.}
        \label{tbl:cylinderExtD4}
\end{table}
\end{landscape}

The set of boundary conditions $(a,m)$ for $D$ and $E$ theories gives
rise to an extended fusion algebra of conformal fields
$\hat\varphi_{(a,m)}$, namely the graph fusion algebra
\begin{equation}
\hat\varphi_{(a,m)} \times\hat\varphi_{(a',m')}=
\sum_{(a'',m'')}\hat
N_{(a,m)\;(a',m')}{}^{(a'',m'')}\hat\varphi_{(a'',m'')}
\end{equation}
with structure constants given by the Verlinde like formula
\begin{equation}
     \hat N_{(a_1,m_1)\;(a_2,m_2)}{}^{(a_3,m_3)}=\sum_{(a',m)}\frac{
     \Psi_{(a_1,m_1)}{}^{(a',m)} \Psi_{(a_2,m_2)}{}^{(a',m)}
     {\Psi_{(a_3,m_3)}{}^{(a',m)}}^*
     }{\Psi_{(1,0)}{}^{(a',m)}}
     \label{eq:Nhat}
\end{equation}
where the sum on $(a',m)\in\text{Exp}(G)\times\Bbb Z_{2k}$ is over the
fundamental domain.  In terms of the extended characters, the cylinder
partition functions can be written as
\begin{equation}
     Z_{(a,m)|(a',m')}(q)=\sum_{(a'',m'')}
     \hat N_{(a,m)^{*}\;(a',m')}{}^{(a'',m'')}\hat\chi_{(a'',m'')}(q)
     \label{eq:Zextended}
\end{equation}
Table~\ref{tbl:cylinderExtD4} gives the example of the $D_{4}$ case. 
In this form the partition functions are no longer explicitly $\Bbb
Z_{3}$ symmetric.  There is a distinguished node which corresponds to
the identity operator of the fusion algebra which for $D_4$ is
$(A,0)$.

\section{Lattice Realizations of ${\Bigbb Z}_k$ Parafermions}
\label{sec:Lattice}

\subsection{$A$-$D$-$E$ lattice models and integrable boundaries}
\label{sec:ADE}
The principal ${\Bbb Z}_k$ theories are realized as the continuum
scaling limit of the critical $A$-$D$-$E$ lattice models~\cite{Pas}
with negative spectral parameter.  In these models the spin states are
taken to be the nodes of the graph $G$ of $A$, $D$ or $E$ type.  Let
$g$ be the Coxeter number of $G$.  Then the bulk face weights are
\setlength{\unitlength}{8mm}
\begin{equation}
\W{a}{b}{c}{d}{u}\;=\;
\begin{picture}(2,0)(0,.85)\multiput(0.5,0.5)(1,0){2}{\line(0,1){1}}
\multiput(0.5,0.5)(0,1){2}{\line(1,0){1}}
\put(0.48,0.63){\pp{bl}{\searrow}}
\put(0.45,0.45){\pp{tr}{a}}\put(1.55,0.45){\pp{tl}{b}}
\put(1.55,1.55){\pp{bl}{c}}\put(0.45,1.55){\pp{br}{d}}
\put(1,1){\pp{}{u}}\put(2.1,0.8){\p{}{}}\end{picture}
\;=
\;\frac{\sin(\lambda-u)}{\sin\lambda}\:\delta_{ac}\;+\;
\frac{\sin u}{\sin\lambda}
\;\frac{\sqrt{\psi_a\:\psi_c}}{\psi_b}\:\delta_{bd}\,,
\end{equation}
where $u$ is the spectral parameter, $\lambda=\pi/g$ is the crossing
parameter and the weights are understood to vanish if the adjacency
condition of $A$, $D$ or $E$ type is not satisfied along the four
edges of the face.  The crossing factors $\psi_a$ are the entries of
the Perron-Frobenius eigenvector of the adjacency matrix $G$.  If
$0<u<\lambda$ the continuum scaling limit of these models describes
the $s\ell(2)$ unitary minimal models.  Otherwise, if
$\lambda-\pi/2<u<0$, the continuum scaling limit describes the
principal ${\Bbb Z}_k$ parafermions with $k=g-2$.

The $A$-$D$-$E$ models are also integrable in the presence of a
boundary~\cite{BPO'B96}.  The integrable boundary conditions at a
conformal point are labelled~\cite{BehrendP} by $(r,a)\in
(A_{g-2},G)$.  A general expression for the boundary weights of the
$(r,a)$ boundary condition~\cite{BehrendP} in terms of boundary edge
weights is \setlength{\unitlength}{6mm}
\begin{equation}
\ba{l}
\ru{3.5}\ds\B{ra}{b}{\beta}{c}{d}{\delta}{u,\xi}\;\;=\;\;
\begin{picture}(2.5,0.0)(0,1.4)
\multiput(2,0.5)(0,1.9){2}{\line(0,1){0.1}}
\multiput(2,0.65)(0,0.25){7}{\line(0,1){0.2}}
\multiput(1.5,0.5)(0,2){2}{\line(1,0){0.5}}
\put(0.5,1.5){\line(1,-1){1}}\put(0.5,1.5){\line(1,1){1}}
\put(1.5,1.7){\pp{}{r,\,a}}\put(1.5,1.3){\pp{}{u,\,\xi}}
\put(1.45,0.47){\pp{tr}{b}}\put(0.4,1.5){\pp{r}{c}}
\put(1.45,2.53){\pp{br}{d}}
\put(2.1,0.5){\pp{tl}{\beta}}\put(2.1,2.5){\pp{bl}{\delta}}
\end{picture}
\\[17pt]
\ds
\mbox{}=\;\frac{\sin(\xi-u)\:\sin(\xi+u+r\lambda)\;\psi_c^\hf}
{\sin(2\xi)\;\psi_{b}^\hf}\;
\delta_{bd}\;\delta_{\beta\delta}\;+\;\frac{\sin(2u)}{\sin(2\xi)}\;
\sum_{\gamma=1}^{F_{ca}^{r+1}}
E^{ra}(b,c)_{\beta\gamma}\:E^{ra}(d,c)_{\delta\gamma}\,.
\ea
\end{equation}
Here $\xi$ is a free parameter that should be thought of as a boundary
field and $E^{ra}(b,c)_{\beta\gamma}$ are edge weights specified in
Behrend and Pearce~\cite{BehrendP}.  The fused adjacency matrices
$F^r$ at level $r$ which appear in the summation are simply another
notation for the intertwiners $F^r=V_{r}^{G}$ defined previously. 
These are given recursively in terms of the adjacency matrix $G$ of
the $A$-$D$-$E$ graph by the $s\ell(2)$ fusion rules
\begin{eqnarray}
F^1=I,\qquad F^2=G,\qquad F^r=GF^{r-1}-F^{r-2},\quad r=3,\ldots,g
\end{eqnarray}
The labels $\alpha,\beta,\gamma$ are bond variables or degeneracies
with $\beta=1,\ldots,F^r_{ab}$, $\delta=1,\ldots,F^r_{ad}$.  At
$u=\xi$ the boundary weights are independent of $\xi$ and decompose
simply in terms of boundary edge weights
\begin{equation}
\B{ra}{b}{\beta}{c}{d}{\delta}{\xi,\xi}\;=\;
\ds\sum_{\gamma=1}^{F_{ca}^{r+1}}E^{ra}(b,c)_{\beta\gamma}\:E^{ra}(d,c
)_{\delta\gamma}
\end{equation}
It is precisely at $u=\xi$, when the triangle boundary weights reduce
effectively to edge weights, that the integrable boundary conditions
correspond~\cite{BehrendP} to conformal boundary conditions.

The $A$-$D$-$E$ face weights and boundary weights satisfy the
Yang-Baxter and boundary Yang-Baxter equations.  This ensures
integrability through commuting double row transfer matrices.  The
entries of the $N$ column double-row transfer matrix with boundary
conditions $(r_1,a_1)$ on the left and $(r_2,a_2)$ on the right are
given by \setlength{\unitlength}{9mm}
\begin{equation}
\ba{l}\ru{3.5}\vec{D}_{\!r_1\!a_1\!|r_2a_2}\!
(u,\xi_1,\xi_2)_{(\beta_1,b_0,\ldots,b_N,\beta_2),(\delta_1,d_0,\ldots
,d_N,\delta_2)}\\
\begin{picture}(9,3)
\put(-0.6,1.37){$=$}
\multiput(0.5,0.5)(0,2){2}{\line(1,0){0.5}}
\multiput(8,0.5)(0,2){2}{\line(1,0){0.5}}
\multiput(2,1.5)(6,-1){2}{\line(-1,1){1}}
\multiput(2,1.5)(6,1){2}{\line(-1,-1){1}}
\multiput(2,0.5)(0,1){3}{\line(1,0){5}}
\multiput(2,0.5)(1,0){2}{\line(0,1){2}}
\multiput(6,0.5)(1,0){2}{\line(0,1){2}}
\multiput(0.5,0.5)(0,0.3){7}{\line(0,1){0.2}}
\multiput(8.5,0.5)(0,0.3){7}{\line(0,1){0.2}}
\multiput(1.99,0.6)(0,1){2}{\pp{bl}{\searrow}}
\multiput(5.99,0.6)(0,1){2}{\pp{bl}{\searrow}}
\multiput(2.5,1)(4,0){2}{\pp{}{u}}
\multiput(2.5,2)(4,0){2}{\pp{}{\lambda-u}}
\put(1.1,1.65){\pp{}{r_1,\,a_1}}\put(1.15,1.35){\pp{}{\lambda-u,\,\xi_1}}
\put(7.9,1.65){\pp{}{r_2,\,a_2}}\put(7.9,1.35){\pp{}{u,\,\xi_2}}
\put(0.46,0.46){\pp{tr}{\beta_1}}\put(0.46,2.54){\pp{br}{\delta_1}}
\put(8.54,0.46){\pp{tl}{\beta_2}}\put(8.54,2.54){\pp{bl}{\delta_2}}
\multiput(1.1,0.39)(0.9,0){2}{\pp{t}{b_0}}\put(3,0.39){\pp{t}{b_1}}
\put(6.1,0.39){\pp{t}{b_{N-1}}}\multiput(7,0.39)(1,0){2}{\pp{t}{b_{N}}}
\multiput(1.1,2.61)(0.9,0){2}{\pp{b}{d_0}}\put(3,2.61){\pp{b}{d_1}}
\put(6.1,2.61){\pp{b}{d_{N-1}}}\multiput(7,2.61)(1,0){2}{\pp{b}{d_{N}}}
\multiput(2,1.5)(1,0){2}{\pp{}{\bullet}}
\multiput(6,1.5)(1,0){2}{\pp{}{\bullet}}
\multiput(1.125,0.5)(0.15,0){6}{\pp{}{.}}
\multiput(1.125,2.5)(0.15,0){6}{\pp{}{.}}
\multiput(7.125,0.5)(0.15,0){6}{\pp{}{.}}
\multiput(7.125,2.5)(0.15,0){6}{\pp{}{.}}
\put(8.9,1.4){\p{}{}}
\end{picture}
\ea
\end{equation}

\subsection{Finite-size corrections} \label{sec:FiniteSize}
The properties of the $A$-$D$-$E$ lattice models connect to the data
of the associated conformal field theories through the finite-size
corrections to the eigenvalues of the double row transfer matrices. 
If we write the eigenvalues of $\vec{D}_{\!r_1\!a_1\!|r_2a_2}\! 
(u,\xi_1,\xi_2)$ as
\begin{equation}
D_n(u)=\exp(-E_n(u)),\quad n=0,1,2,\ldots
\end{equation}
then the finite-size corrections to the energies $E_n$ take the form
\begin{equation}
E_n(u)=2Nf(u)+f_{r_1,a_1|r_2,a_2}(u)
+\frac{2\pi\sin\vartheta}{N}\,\left(-{c\over 24}+\Delta_n+k_n\right)
+o\left(\frac{1}{N}\right),
\;\; k_n\in{\Bbb N}
\end{equation}
where $f(u)$ is the bulk free energy, $f_{r_1,a_1|r_2,a_2}(u)$ is the
boundary free energy, $c$ is the central charge, $\Delta_n$ is a
conformal weight and the anisotropy angle is given by
\begin{equation}
\vartheta=\cases{
g u,& \text{$0<u<\lambda$\qquad\qquad (unitary minimal)}\\[.5cm]
-\frac{2(k+2)}{k}\, u,&\text{$\lambda-\pi/2<u<0$\quad \,(${\Bbb Z}_k$
parafermions)}}
\end{equation}
where $g=k+2$ is the Coxeter number.

Removing the bulk and boundary contributions to the partition function
on a cylinder leads to the conformal partition function $Z_{j|k}(q)$
with left and right boundaries $j=(r_1,a_1)$, $k=(r_2,a_2)$.  This can
be expressed as a linear form in characters
\begin{equation}
Z_{j|k}(q)=\sum_i n_{ij}{}^k \chi_i(q)
\end{equation}
where $i$ is summed over the primary fields and the integers
$n_{ij}{}^k\in{\Bbb N}$ give the operator content.  For $M$ double
rows the modular parameter is
\begin{equation}
q=\exp(2\pi i\tau),\qquad \tau=i\frac{M}{N} \sin\vartheta
\end{equation}
where $M/N$ is the aspect ratio of the cylinder.

For a system to be conformally invariant it is usually demanded that
it is both isotropic and translationally invariant.  For the unitary
minimal $A$-$D$-$E$ models with $0<u<\lambda$ the geometry is that of
an isotropic square lattice when $u=\lambda/2$ and $\vartheta=\pi/2$. 
In this case the alternation of rows in the double row transfer matrix
also disappears since $\lambda-u=u$.  So the conformal point occurs
for
\begin{equation}
\lambda-u=u=\xi_1=\xi_2=\lambda/2\qquad\text{(unitary minimal models)}
\end{equation}
In contrast, for the ${\Bbb Z}_k$ parafermionic $A$-$D$-$E$ models
with $\lambda-\pi/2<u<0$, the relevant geometry is not that of an
isotropic square lattice.  Instead, for ${\Bbb Z}_k$ parafermions, the
relevant choice is $u=-\lambda$ with $\vartheta=2\pi/k$.  Thus for the
${W}_3$ or hard hexagon model the geometry is that of the
triangular lattice with $\vartheta=2\pi/3$.  In general it follows
that $\lambda-u=2\lambda$ so the alternation of rows in the double row
transfer matrix persists even though the second row is still the
transpose of the first row.  Moreover, to ensure that the left and
right boundary conditions are conformal we need to choose
$\xi_1=2\lambda$ and $\xi_2=-\lambda$ so the conformal point is
\begin{equation}
\lambda-u=\xi_1=2\lambda,\qquad u=\xi_2=-\lambda\qquad \text{(${\Bbb
Z}_k$ parafermions)}\label{chiral}
\end{equation}
Notice that at this conformal point the double row transfer matrix is
not left-right symmetric and the boundary free energies for a given
boundary condition $(r,a)$ on the left and right are different because
they have different boundary fields $\xi_1=2\lambda$ and
$\xi_2=-\lambda$.

\subsection{Bulk free energies} \label{sec:Bulk}

In this section we obtain the bulk free energies $f(u)$ or
equivalently the partition functions per site $\kappa(u)=\exp(-f(u))$. 
Two $A$-$D$-$E$ models sharing the same Coxeter number are related by
intertwiners so their bulk free energies are the same.  Their boundary
free energies are also related.  Thus we only need to find the free
energies for the $A_{L}$ models.

For general $L=k+1$, the partition function per site of the $A_L$
model satisfies the {\em inversion relation}
\begin{equation}
\kappa (u)\; \kappa
(u+\lambda)=\frac{\sin(u+\lambda)\sin(u-\lambda)}{\sin^2\lambda}
\frac{\sin\frac{L+1}{L-1}\,u}{\sin\frac{L+1}{L-1}(u+\lambda)}:=q_b(u)
        \label{eq:inversion}
\end{equation}
and the {\em crossing symmetry}
\begin{equation}
       \kappa (u)=\kappa (\lambda -u).
        \label{eq:crossing}
\end{equation}
The inversion relation is established by keeping just the dominant
terms in the TBA and inversion identity hierarchies~\cite{BPO'B96},
using height reversal symmetry and equating just the bulk terms of
order $2N$.  The inversion relation, crossing symmetry and height
reversal symmetry determine a unique solution which is analytic and
non-zero in the analyticity strip $\Re(u)\in(\frac{-\pi+ \lambda}
2,\frac\lambda 2)$, which contains the physical strip
$\Re(u)\in(-\frac{\pi}2+ \lambda,0)$ for the regime we consider.  This
problem has been solved, in a different context, by Baxter~\cite{Bax}. 
The solution for general $L$ can be written as an integral
\begin{equation}
       \kappa (u)=\exp\int_{-\infty}^{+\infty}\sinh u t\;
        \frac{
        \sinh (\pi-\lambda)t\;
        \sinh (\pi-3\lambda+u)t-\sinh\lambda t\;
        \sinh (u+\lambda)t}
        {t\sinh\pi t\;
        \sinh (\pi-2\lambda)t}\;
        dt.
        \label{eq:eq:fLodd}
\end{equation}
The solution can also be written explicitly for $L$ even ($k$ odd) as
\begin{equation}
\kappa (u)=\frac{\sin(u-2\lambda)}{\sin\lambda}\prod_{k=1}^{L/2-1}
\frac{\sin(u+(2k-1)\lambda)}{\sin(u+2k\lambda)}
\frac{\sin\frac{L+1}{L-1}(u+2k\lambda)}{\sin\frac{L+1}{L-1}(u+(2k-1)\lambda)}.
        \label{eq:fLeven}
\end{equation}

%

\subsection{Boundary free energies} \label{sec:Boundary}

In this section we obtain results for the boundary free energies by
extending the inversion relation method~\cite{O'BPB96,O'BP}.  The
boundary free energy contribution comes in three different parts
\begin{equation}
        f_{r,s|r',s'}=f_{0}+f_{r,s}^{L}+f_{r',s'}^{R}.
        \label{eq:frs}
\end{equation}
The vacuum term $f_{0}$ does not depend on the boundary condition.  It
is free of zeros and its logarithm is analytic on the physical
analyticity strip.  The remaining contributions $f_{r,s}^{L/R}$ for
each boundary condition $(r,s)$ are different on the left $L$ and on
the right $R$ and exhibit zeros in the physical analyticity strip. 
The boundary free energies are obtained by repeating the analysis of
the TBA and inversion identity hierarchies~\cite{BPO'B96} keeping the
dominant terms as in the bulk calculation but this time equating the
boundary terms of order 1 in the physical analyticity strip
$\Re(u)\in(\frac{-\pi +\lambda} 2,-\frac\lambda 2)$.

Let $q_b$ be the RHS of the bulk inversion relation
(\ref{eq:inversion}).  Then the vacuum functional equation we obtain
is, with $\kappa_{0}(u)=\exp(-f_{0}(u))$,
\begin{equation}
\kappa_{0}(u)\; \kappa_{0}(u+\lambda)=q_b(u)\;  q_b(-\frac\pi
2-u)\;\frac{4\sin^4
\lambda}{\sin (\lambda-2u)\sin (\lambda-2(u+\lambda))}.
      \label{eq:febound}
\end{equation}
Hence we find the solution with the required analyticity is expressed
in terms of the bulk free energy as
\begin{equation}
\kappa _{0}(u)=\kappa (u)\; \kappa (-\frac\pi 2-u+\lambda)\; \frac{2\,\sin^2
\lambda}{\sin
(\lambda-2u)}
\end{equation}
For $L$ even, one has an explicit expression
\begin{eqnarray}
\kappa _{0}(u)&=&\frac{\sin 2(u+\lambda)}{\sin (\lambda-2u)} \\ &&
\times\prod_{k=1}^{\frac{L}2 -1}
\frac{\sin 2(u+(2k +1)\lambda)}{\sin 2(u+2k\lambda)}
\frac{\sin \frac{L+1}{L-1}(u+2k\lambda)}{\sin
\frac{L+1}{L-1}(u+(2k -1)\lambda)}
\frac{\cos
\frac{L+1}{L-1}(u+(2k -1)\lambda)}{\cos \frac{L+1}{L-1}(u+2k\lambda)}
\notag
\end{eqnarray}
In this case the inversion relation admits other explicit solutions. 
Combining the zeros and poles of the two bulk free energy terms for
$L\equiv 1 \mathrm{~mod~} 4$, we find
\begin{eqnarray}
\kappa _{0}(u)&=&\frac{\sin 2(u+(\frac{L-1}2 -1)\lambda)}{\sin (2u
-\lambda)}
\frac{\sin \frac{L+1}{L-1}(u-\lambda)}{\sin
\frac{L+1}{L-1}(u+(\frac{L-1}2 -1)\lambda)}\\ &&
\times\prod_{k=1}^{\frac{L-1}4 -1}
\frac{\sin 2(u+(2k -1)\lambda)}{\sin 2(u+2k\lambda)}
\frac{\sin \frac{L+1}{L-1}(u+2k\lambda)}{\sin
\frac{L+1}{L-1}(u+(2k -1)\lambda)}
\frac{\cos
\frac{L+1}{L-1}(u+(2k -1)\lambda)}{\cos \frac{L+1}{L-1}(u+2k\lambda)}
\notag
\end{eqnarray}

In the remaining case $L\equiv 3 \mathrm{~mod~} 4$, the vacuum
boundary free energy can only be expressed in terms of integrals.  Let
$q_{0}$ be the RHS of the functional equation~(\ref{eq:febound}).  It
is Analytic and Non-Zero in the strip $\Re(u)\in(\frac{-\pi+ \lambda}
2,-\frac \lambda 2)$.  Furthermore the derivative $f'_{0}$ approaches
a Constant when $\Im(u)\to\pm\infty$ (ANZC).  Hence we can introduce
the Fourier transforms of the derivatives
\begin{eqnarray}
\mathcal{F}_0(k)&:=&{\frac 1{2 i
\pi}}\int\limits_{\hidewidth\frac{-\pi+
\lambda} 2<\Re(u)<-\frac \lambda 2\hidewidth}f'_{0}(u)e^{-ku}du \\
{\frac d{du}}f_{0}(u)&=&
\int\limits_{-\infty}^{+\infty}\mathcal{F}_0(k)e^{ku}dk
\end{eqnarray}
so that (\ref{eq:febound}) becomes
\begin{equation}
\mathcal{F}_0(k)(1+e^{k\lambda})={\frac 1{2 i
\pi}}\int\limits_{\hidewidth\frac{-\pi+ \lambda} 2<\Re(u)<-\frac \lambda
2\hidewidth}\frac{q'_0(u)}{q_0(u)}e^{-ku}du
      \label{eq:Ffebound}
\end{equation}
The solution by inverse Fourier transform gives
\begin{eqnarray}
{\frac d{du}}f_0(u)&=&{\frac 1{2 i\pi}}
\int\limits_{\hidewidth0<\Re(w)<\min (\lambda,u+\frac L2\lambda)\hidewidth}d
w\left({\frac d{du}}\;
\log q_0(u-w)\right)\frac{L+1}{\sin (L+1)w}
      \label{eq:soldbound}
\end{eqnarray}
Integrating with respect to $u$ and taking the $w$ integration along
the vertical line $\Re(w)=\epsilon>0$ we obtain in the limit
$\epsilon\to 0$
\begin{eqnarray}
f_0(u)
&=&\frac{\log q_0(u)}{2}
+\frac{L+1}{\pi}\int_0^\infty
\frac {\text{Im } \log q_0(u-iw)}{\sinh (L+1)w}d w
      \label{eq:solbound}
\end{eqnarray}

The boundary free energy $f_{(r,s)}$ arises from a boundary condition
of type $(r,s)$ with a boundary field $\xi$.  It does not depend on
$s$.  An integral form analogous to (\ref{eq:solbound}) can be derived
with $q_{0}$ being replaced by the RHS of the boundary inversion
relation
\begin{eqnarray}
\kappa _{(r,s)}(u)\; \kappa _{(r,s)}(u+\lambda)&=&
\frac{\sin(u+\xi)\sin(u-\xi)\sin(u+\xi+r\lambda)\sin(u-\xi-r\lambda)}
{\sin^4\lambda}\notag\\
&&
\,\times\frac{\sin\frac{L+1}{L-1}(u-\xi-(r-1)\lambda)}{\sin\frac{L+1}{
L-1}(u-\xi)}
\,\frac{\sin\frac{L+1}{L-1}(u+\xi+\lambda)}{\sin\frac{L+1}{L-1}(u+\xi+
r\lambda)}
\end{eqnarray}
The solution must be analytic on the analyticity strip with the same
zeros as
\begin{equation}
\sin(u+\xi)\sin(u-\xi-r\lambda).
\end{equation}
For $r$ odd, this inversion relation admits the explicit solution
\begin{eqnarray}
\kappa _{(r,s)}(u)&=&\frac{\sin (u+\xi)\sin (u-\xi-r\lambda)}
{\sin^2\lambda}\\ && \times\prod_{k=1}^{\frac{r-1}2} \,\frac{\sin
(u+\xi+2k\lambda)} {\sin (u+\xi+(2k-1)\lambda)} \frac{\sin
(u-\xi-(2k-1)\lambda)}{\sin (u-\xi-2k\lambda)} \notag\\ &&
\,\qquad\times\frac{
\sin\frac{L+1}{L-1}(u+\xi+(2k-1)\lambda)}{\sin\frac{L+1}{L-1}(u+\xi+2k\lambda)}
\,\frac{
\sin\frac{L+1}{L-1}(u-\xi-2k\lambda)}{\sin\frac{L+1}{L-1}(u-\xi-(2k-1)\lambda)}
\notag\end{eqnarray} with zeros at $u=-\xi,\, \xi+r\lambda,\,
-\xi-\pi+\lambda,\, \xi-\pi+(r+1)\lambda$.  Since these last two zeros
are not desired, the formula is only valid when these points are
outside the analyticity strip, {\it i.e.},
\begin{equation}
\frac{-\pi+\lambda}2<\Re(\xi)<\frac{\pi-\lambda}2 -r\lambda.
\end{equation}
%
%
There is another solution when $L+r$ is even
\begin{eqnarray}
\kappa _{(r,s)}(u)&=&\frac{\sin (u+\xi-\lambda)\sin (u-\xi)}
{\sin^2\lambda}
\\ &&
\times\prod_{k=1}^{\frac{L-r}2}
\,\frac{\sin (u+\xi+(2(k-1)+r)\lambda)}
{\sin (u+\xi+(2k-1+r)\lambda)}
\frac{\sin (u-\xi+2k\lambda)}{\sin (u-\xi+(2k-1)\lambda)}
\notag\\ &&
\,\qquad\times\frac{
\sin\frac{L+1}{L-1}(u+\xi+(2k-1+r)\lambda)}{\sin\frac{L+1}{L-1}(u+\xi+
(2(k-1)+r)\lambda)}
\,\frac{
\sin\frac{L+1}{L-1}(u-\xi+(2k-1)\lambda)}{\sin\frac{L+1}{L-1}(u-\xi+2(
k-1)\lambda)}
\notag\end{eqnarray}
with zeros at $u=-\xi, \xi+r\lambda,
\xi+(r-1)\lambda-\pi,-\xi-(r-2)\lambda-\pi$. Clearly, this formula is
only valid when
\begin{equation}
\frac{-\pi+3\lambda}2-r\lambda<\Re(\xi)<\frac{\pi-\lambda}2 -r\lambda.
\end{equation}

In the $A_4$ and $A_5$ cases, these formulas allow us to compute the
boundary free energies.  The values for $A_{5}$ are listed in
Table~\ref{tbl:freeA5}.
\begin{table}[htbp]
        \centering
        $$ f_{0}(-\lambda/2) =\log \frac{1+\sqrt{2}}3 $$
        $$\begin{array}{rcccc}
        r\ =& 1\quad&\quad 2 \quad\mbox{}&\quad 3\quad\mbox{}&\quad
4\quad\mbox{}\\[8pt]
        f_{(r,s)}(-\lambda/2)\ =&0&
         \half\log \frac83 & \half\log 3 & \half\log \frac32\\[8pt]
      f_{(r,s)}(3\lambda/2)\ = &0& \half\log 2 & \log \frac43 & \half\log
\frac23
     \end{array}$$
        \caption{The right and left boundary free energies
$f_{(r,s)}^R=f_{(r,s)}(u)$ and
$f_{(r,s)}^L=f_{(r,s)}(\lambda-u)$ of the $A_{5}$ model at the special point
$u=-\lambda/2$ and
$\lambda-u=3\lambda/2$. The boundary free energies are independent of $s$.}
        \label{tbl:freeA5}
\end{table}

Some symmetries of the parafermion models are reflected in the
boundary free energies.  The left/right symmetry is broken in general
but the usual height reversal symmetry $r,s\mapsto r,L+1-s$ remains. 
Another symmetry comes from the fact that at a conformal point the
boundary weights factorize into edge weights.  The interchange of the
two independent sublattices is then implemented by the Kac table
symmetry $r,s\mapsto L-r,L+1-s$.

\section{Identifying integrable and conformal boundary conditions}
\label{sec:Identify} For the unitary minimal $A$-$D$-$E$ models with
$0<u<\lambda$ there is a one-to-one correspondence~\cite{BehrendP}
between integrable and conformal boundary conditions labelled by
$(r,a)\in (A_{g-2},G)$ that respect the symmetry of the generalized
Kac table.  In contrast, for the ${\Bbb Z}_k$ parafermionic
$A$-$D$-$E$ models with $\lambda-\pi/2<u<0$, the situation is more
subtle and the conformal labels $(a,m)\in(G,{\Bbb Z}_{2k})$ do not
coincide with the natural labels $(r,a)$ arising from the
construction~\cite{BehrendP} of integrable boundary conditions at the
conformal point.  It is therefore necessary to make an identification
of the integrable and conformal boundary conditions from numerical
data.

For the $A$ type ${\Bbb Z}_k$ models the dominant or vacuum
configuration is given by the $2k$ periodic sawtooth configuration
\begin{equation}
\vec a=\{\ldots,1,2,\ldots,k,k+1,k,\ldots,2,1,\ldots\}
\end{equation}
It is thus clear that the vacuum $(\ell,m)=(0,0)$ on the left and the
right should be identified with the integrable boundary condition
$(r,a)=(1,1)$ on the left and the right with the number of faces $N=0$
mod $2k$ and this is confirmed by numerical computation.  Next, if we
fix the boundary condition to the vacuum $(\ell,m)=(0,0)$ on the left
or right we find from (\ref{Verlinde}) that the cylinder partition
function reduces to a single character
\begin{equation}
Z_{(0,0)|(\ell,m)}(q)=Z_{(\ell,m)|(0,0)}(q)=\chi_{(\ell,m)}(q)
\label{singchar}
\end{equation}
If we fix $N$ mod $2k$, this allows us to identify, up to possible
ambiguities of conjugation, the integrable boundary condition $(r,s)$
on the left or right with the conformal labels $(\ell,m)$ from the
numerically determined cylinder partition functions.  Once these
identifications are made they can be checked for consistency against
all the other cylinder partition functions given in (\ref{Verlinde}).

Proceeding in this way we find the following correspondence for the
$A_{L}$ parafermions with $N=0$ mod $2k$
\begin{equation}
(\ell,m)=
\cases{(s-1,\; r-1),&\mbox{$r+s$ even}\\
(k+1-s,\; k-r),&\mbox{$r+s$ odd.}
}
\label{eq:correspondence}
\end{equation}
This is consistent with the Kac table symmetries, $(r,s)\mapsto
(k+1-r,k+2-s)$ and (\ref{eq:sym}), of both the minimal models and
parafermions.  The correspondence for the $A_3$, $A_4$ and $A_5$
parafermion models is listed explicitly in Table~\ref{tbl:corres}. 
The height reversal symmetry $s\mapsto k+2-s$, flips the tables about
their middle rows and corresponds at the level of the conformal
algebra to a fusion with the field $\varphi_{(0,k)}$ which is of order
two.  The conjugacy $(l,m)\mapsto (k-l,k-m)$ can be seen in these
tables as a central symmetry which centre has been indicated {\em for
the even sublattice}.  The centre of the usual Kac table symmetry of
the minimal models has been indicated too.

\begin{table}[htbp]
\nc{\spos}[2]{\makebox(0,0)[#1]{${#2}$}}
\nc{\sm}[1]{{\scriptstyle #1}}
\setlength{\unitlength}{9mm}
\begin{center}
\vspace{1cm}
      \begin{picture}(5,3)
\put(3,4.2){\spos{}{A_3\quad (k=2)}}
\multiput(0,0)(1,0){3}{\line(0,1){3}}
\multiput(0,0)(0,1){4}{\line(1,0){2}}
\put(1.0 , 1.5){\spos{}{\bullet}}
\put(-1,.5){\line(0,1){2}}
\multiput(-1,.5)(0,1){3}{\spos{}{\bullet}}
\put(.5,-1){\line(1,0){1}}
\multiput(.5,-1)(1,0){2}{\spos{}{\bullet}}
\put(0.5,0.5){\spos{}{\scriptstyle (0,0)}}
\put(1.5,0.5){\spos{}{\scriptstyle (0,2)}}
\put(0.5,1.5){\spos{}{\scriptstyle (1,1)}}
\put(1.5,1.5){\spos{}{\scriptstyle (1,1)}}
\put(0.5,2.5){\spos{}{\scriptstyle (0,2)}}
\put(1.5,2.5){\spos{}{\scriptstyle (0,0)}}
\put(0.5,-0.5){\spos{}{1}}
\put(1.5,-0.5){\spos{}{2}}
\put(3,-0.5){\spos{}{r\in A_{2}}}
\put(-0.5,0.5){\spos{}{1}}
\put(-0.5,1.5){\spos{}{2}}
\put(-0.5,2.5){\spos{}{3}}
\put(-0.5,3.5){\spos{}{s\in A_{3}}}
\end{picture}
      \begin{picture}(4,3)
\multiput(0,0)(1,0){5}{\line(0,1){3}}
\multiput(0,0)(0,1){4}{\line(1,0){4}}
\put(2.02,0){\line(0,1){3}}
\put(-1,.5){\line(0,1){2}}
\multiput(-1,.5)(0,1){3}{\spos{}{\bullet}}
\put(-.3,-1){\line(1,0){4.6}}
\multiput(.5,-1)(1,0){4}{\spos{}{\bullet}}
\multiput(-.05,-1)(1,0){5}{\spos{}{>}}
\put(0.5,0.5){\spos{}{\topped{(1,1)}{(2,3)}}}
\put(2.5,0.5){\spos{}{\topped {(1,3)}{(2,1)}}}
\put(1.5,1.5){\spos{}{\topped {(2,2)}{(1,2)}}}
\put(3.5,1.5){\spos{}{\topped {(2,2)}{(1,2)}}}
\put(0.5,2.5){\spos{}{\topped {(1,3)}{(2,1)}}}
\put(2.5,2.5){\spos{}{\topped {(1,1)}{(2,3)}}}
\put(0.5,-0.5){\spos{}{0}}
\put(1.5,-0.5){\spos{}{1}}
\put(2.5,-0.5){\spos{}{2}}
\put(3.5,-0.5){\spos{}{3}}
\put(4.5,-0.5){\spos{}{m\in \Bbb Z_{4}}}
\put(-0.5,0.5){\spos{}{0}}
\put(-0.5,1.5){\spos{}{1}}
\put(-0.5,2.5){\spos{}{2}}
\put(-0.5,3.5){\spos{}{\ell\in A_{3}}}
\end{picture}
\vspace{0.35in}
\end{center}
\begin{center}
\vspace{1.5cm}
\begin{picture}(6,4)
\put(5.0,5.2){\spos{}{A_4\quad (k=3)}}
\multiput(0,0)(1,0){4}{\line(0,1){4}}
\multiput(0,0)(0,1){5}{\line(1,0){3}}
\put(2,2){\spos{}{*}}
\put(1.5,2){\spos{}{\bullet}}
\put(-1,.5){\line(0,1){3}}
\multiput(-1,.5)(0,1){4}{\spos{}{\bullet}}
\put(0.5,-1){\line(1,0){2}}
\multiput(.5,-1)(1,0){3}{\spos{}{\bullet}}
\put(0.5,0.5){\spos{}{\scriptstyle (0,0)}}
\put(1.5,0.5){\spos{}{\scriptstyle (3,1)}}
\put(2.5,0.5){\spos{}{\scriptstyle (0,2)}}
\put(0.5,1.5){\spos{}{\scriptstyle (2,2)}}
\put(1.5,1.5){\spos{}{\scriptstyle (1,1)}}
\put(2.5,1.5){\spos{}{\scriptstyle (2,0)}}
\put(0.5,2.5){\spos{}{\scriptstyle (2,0)}}
\put(1.5,2.5){\spos{}{\scriptstyle (1,1)}}
\put(2.5,2.5){\spos{}{\scriptstyle (2,2)}}
\put(0.5,3.5){\spos{}{\scriptstyle (0,2)}}
\put(1.5,3.5){\spos{}{\scriptstyle (3,1)}}
\put(2.5,3.5){\spos{}{\scriptstyle (0,0)}}
\put(0.5,-0.5){\spos{}{1}}
\put(1.5,-0.5){\spos{}{2}}
\put(2.5,-0.5){\spos{}{3}}
\put(3.5,-0.5){\spos{}{r}}
\put(-0.5,0.5){\spos{}{1}}
\put(-0.5,1.5){\spos{}{2}}
\put(-0.5,2.5){\spos{}{3}}
\put(-0.5,3.5){\spos{}{4}}
\put(-0.5,4.5){\spos{}{s}}
\end{picture}
\begin{picture}(6,4)
\multiput(0,0)(1,0){7}{\line(0,1){4}}
\multiput(0,0)(0,1){5}{\line(1,0){6}}
\put(3.02,0){\line(0,1){4}}
\put(2,2){\spos{}{*}}
\put(-1,.5){\line(0,1){3}}
\multiput(-1,.5)(0,1){4}{\spos{}{\bullet}}
\put(-.3,-1){\line(1,0){6.6}}
\multiput(.5,-1)(1,0){6}{\spos{}{\bullet}}
\multiput(-.05,-1)(1,0){7}{\spos{}{>}}
\put(0.5,0.5){\spos{}{\topped{(1,1)}{(3,4)}}}
\put(2.5,0.5){\spos{}{\topped{(3,1)}{(1,4)}}}
\put(4.5,0.5){\spos{}{\topped{(2,4)}{(2,1)}}}
\put(1.5,1.5){\spos{}{\topped{(2,2)}{(2,3)}}}
\put(3.5,1.5){\spos{}{\topped{(1,3)}{(3,2)}}}
\put(5.5,1.5){\spos{}{\topped{(3,3)}{(1,2)}}}
\put(0.5,2.5){\spos{}{\topped{(1,3)}{(3,2)}}}
\put(2.5,2.5){\spos{}{\topped{(3,3)}{(1,2)}}}
\put(4.5,2.5){\spos{}{\topped{(2,3)}{(2,2)}}}
\put(1.5,3.5){\spos{}{\topped{(2,4)}{(2,1)}}}
\put(3.5,3.5){\spos{}{\topped{(1,1)}{(3,4)}}}
\put(5.5,3.5){\spos{}{\topped{(3,1)}{(1,4)}}}
\put(0.5,-0.5){\spos{}{0}}
\put(1.5,-0.5){\spos{}{1}}
\put(2.5,-0.5){\spos{}{2}}
\put(3.5,-0.5){\spos{}{3}}
\put(4.5,-0.5){\spos{}{4}}
\put(5.5,-0.5){\spos{}{5}}
\put(6.5,-0.5){\spos{}{m}}
\put(-0.5,0.5){\spos{}{0}}
\put(-0.5,1.5){\spos{}{1}}
\put(-0.5,2.5){\spos{}{2}}
\put(-0.5,3.5){\spos{}{3}}
\put(-0.5,4.5){\spos{}{\ell}}
\end{picture}
\vspace{1.5cm}
\end{center}
\begin{center}
\vspace{0.55in}
\begin{picture}(6,5)
\put(6.0,6.2){\spos{}{A_5\quad (k=4)}}
\multiput(0,0)(1,0){5}{\line(0,1){5}}
\multiput(0,0)(0,1){6}{\line(1,0){4}}
\put(2.5,2.5){\spos{}{*}}
\put(2,2.5){\spos{}{\bullet}}
\put(-1,.5){\line(0,1){4}}
\multiput(-1,.5)(0,1){5}{\spos{}{\bullet}}
\put(.5,-1){\line(1,0){3}}
\multiput(.5,-1)(1,0){4}{\spos{}{\bullet}}
\put(0.5,0.5){\spos{}{\scriptstyle (0,0)}}
\put(1.5,0.5){\spos{}{\scriptstyle (4,2)}}
\put(2.5,0.5){\spos{}{\scriptstyle (0,2)}}
\put(3.5,0.5){\spos{}{\scriptstyle (4,0)}}
\put(0.5,1.5){\spos{}{\scriptstyle (3,3)}}
\put(1.5,1.5){\spos{}{\scriptstyle (1,1)}}
\put(2.5,1.5){\spos{}{\scriptstyle (3,1)}}
\put(3.5,1.5){\spos{}{\scriptstyle (1,3)}}
\put(0.5,2.5){\spos{}{\scriptstyle (2,0)}}
\put(1.5,2.5){\spos{}{\scriptstyle (2,2)}}
\put(2.5,2.5){\spos{}{\scriptstyle (2 \;\; 2)}}
\put(3.5,2.5){\spos{}{\scriptstyle (2,0)}}
\put(0.5,3.5){\spos{}{\scriptstyle (1,3)}}
\put(1.5,3.5){\spos{}{\scriptstyle (3,1)}}
\put(2.5,3.5){\spos{}{\scriptstyle (1,1)}}
\put(3.5,3.5){\spos{}{\scriptstyle (3,3)}}
\put(0.5,4.5){\spos{}{\scriptstyle (4,0)}}
\put(1.5,4.5){\spos{}{\scriptstyle (0,2)}}
\put(2.5,4.5){\spos{}{\scriptstyle (4,2)}}
\put(3.5,4.5){\spos{}{\scriptstyle (0,0)}}
\put(0.5,-0.5){\spos{}{1}}
\put(1.5,-0.5){\spos{}{2}}
\put(2.5,-0.5){\spos{}{3}}
\put(3.5,-0.5){\spos{}{4}}
\put(4.5,-0.5){\spos{}{r}}
\put(-0.5,0.5){\spos{}{1}}
\put(-0.5,1.5){\spos{}{2}}
\put(-0.5,2.5){\spos{}{3}}
\put(-0.5,3.5){\spos{}{4}}
\put(-0.5,4.5){\spos{}{5}}
\put(-0.5,5.5){\spos{}{s}}
\end{picture}
\begin{picture}(8,5)
\multiput(0,0)(1,0){9}{\line(0,1){5}}
\multiput(0,0)(0,1){6}{\line(1,0){8}}
\put(4.02,0){\line(0,1){5}}
\put(2.5,2.5){\spos{}{*}}
\put(-1,.5){\line(0,1){4}}
\multiput(-1,.5)(0,1){5}{\spos{}{\bullet}}
\put(-.3,-1){\line(1,0){8.6}}
\multiput(.5,-1)(1,0){8}{\spos{}{\bullet}}
\multiput(-.05,-1)(1,0){9}{\spos{}{>}}
\put(0.5,0.5){\spos{}{\topped{(1,1)}{(4,5)}}}
\put(2.5,0.5){\spos{}{\topped{(3,1)}{(2,5)}}}
\put(4.5,0.5){\spos{}{\topped{(1,5)}{(4,1)}}}
\put(6.5,0.5){\spos{}{\topped{(3,5)}{(2,1)}}}
\put(1.5,1.5){\spos{}{\topped{(2,2)}{(3,4)}}}
\put(3.5,1.5){\spos{}{\topped{(4,2)}{(1,4)}}}
\put(5.5,1.5){\spos{}{\topped{(2,4)}{(3,2)}}}
\put(7.5,1.5){\spos{}{\topped{(4,4)}{(1,2)}}}
\put(0.5,2.5){\spos{}{\topped{(1,3)}{(4,3)}}}
\put(2.5,2.5){\spos{}{\topped{(3 \;\; 3)}{(2 \;\; 3)}}}
\put(4.5,2.5){\spos{}{\topped{(1,3)}{(4,3)}}}
\put(6.5,2.5){\spos{}{\topped{(3,3)}{(2,3)}}}
\put(1.5,3.5){\spos{}{\topped{(2,4)}{(3,2)}}}
\put(3.5,3.5){\spos{}{\topped{(4,4)}{(1,2)}}}
\put(5.5,3.5){\spos{}{\topped{(2,2)}{(3,4)}}}
\put(7.5,3.5){\spos{}{\topped{(4,2)}{(1,4)}}}
\put(0.5,4.5){\spos{}{\topped{(1,5)}{(4,1)}}}
\put(2.5,4.5){\spos{}{\topped{(3,5)}{(2,1)}}}
\put(4.5,4.5){\spos{}{\topped{(1,1)}{(4,5)}}}
\put(6.5,4.5){\spos{}{\topped{(3,1)}{(2,5)}}}
\put(0.5,-0.5){\spos{}{0}}
\put(1.5,-0.5){\spos{}{1}}
\put(2.5,-0.5){\spos{}{2}}
\put(3.5,-0.5){\spos{}{3}}
\put(4.5,-0.5){\spos{}{4}}
\put(5.5,-0.5){\spos{}{5}}
\put(6.5,-0.5){\spos{}{6}}
\put(7.5,-0.5){\spos{}{7}}
\put(8.5,-0.5){\spos{}{m}}
\put(-0.5,0.5){\spos{}{0}}
\put(-0.5,1.5){\spos{}{1}}
\put(-0.5,2.5){\spos{}{2}}
\put(-0.5,3.5){\spos{}{3}}
\put(-0.5,4.5){\spos{}{4}}
\put(-0.5,5.5){\spos{}{\ell}}
\end{picture}
\end{center}
	\vspace{1cm} \caption{The correspondence $(r,s)\mapsto
	(\ell,m)$ between the construction labels $(r,s)\in
	A_{k}\times A_{k+1}$ of the integrable boundary conditions and
	the parafermionic conformal fields $(\ell,m)\in A_{k+1}\times
	\Bbb Z_{2k}$ and its inverse, for $k=2, 3, 4$ and $N=0$ mod
	$2k$.  The inverse is a 1:2 mapping.  The conjugation is
	realized in the {\em even sub-lattices} by a central inversion
	symmetry through the point indicated by $*$, and the minimal
	model symmetry centre is indicated by $\bullet$.}
        \label{tbl:corres}
\end{table}

\begin{table}[htbp]
      \centering
\nc{\spos}[2]{\makebox(0,0)[#1]{${#2}$}}
\nc{\sm}[1]{{\scriptstyle #1}}
\setlength{\unitlength}{9mm}
\hbox{\kern9mm
\raise18mm\hbox{
      \begin{picture}(5,3)
\put(1.,4.){\spos{}{A_3\quad (k=2)}}
\multiput(0,0)(1,0){3}{\line(0,1){3}}
\multiput(0,0)(0,1){4}{\line(1,0){2}}
\put(0.5,0.5){\spos{}{\scriptstyle (0,2)}}
\put(1.5,0.5){\spos{}{\scriptstyle (0,2)}}
\put(0.5,1.5){\spos{}{\scriptstyle (1,1)}}
\put(1.5,1.5){\spos{}{\scriptstyle (1,1)}}
\put(0.5,2.5){\spos{}{\scriptstyle (0,0)}}
\put(1.5,2.5){\spos{}{\scriptstyle (0,0)}}
\put(0.5,-0.5){\spos{}{1}}
\put(1.5,-0.5){\spos{}{2}}
\put(2.5,-0.5){\spos{}{r}}
\put(-0.5,0.5){\spos{}{1}}
\put(-0.5,1.5){\spos{}{2}}
\put(-0.5,2.5){\spos{}{3}}
\put(-0.5,3.5){\spos{}{s}}
\end{picture}
}
\raise9mm\hbox{
\begin{picture}(6,4)
\put(1.5,5.){\spos{}{A_4\quad (k=3)}}
\multiput(0,0)(1,0){4}{\line(0,1){4}}
\multiput(0,0)(0,1){5}{\line(1,0){3}}
\put(0.5,0.5){\spos{}{\scriptstyle (3,1)}}
\put(1.5,0.5){\spos{}{\scriptstyle (0,2)}}
\put(2.5,0.5){\spos{}{\scriptstyle (0,2)}}
\put(0.5,1.5){\spos{}{\scriptstyle (1,1)}}
\put(1.5,1.5){\spos{}{\scriptstyle (2,0)}}
\put(2.5,1.5){\spos{}{\scriptstyle (2,0)}}
\put(0.5,2.5){\spos{}{\scriptstyle (1,1)}}
\put(1.5,2.5){\spos{}{\scriptstyle (2,2)}}
\put(2.5,2.5){\spos{}{\scriptstyle (2,2)}}
\put(0.5,3.5){\spos{}{\scriptstyle (3,1)}}
\put(1.5,3.5){\spos{}{\scriptstyle (0,0)}}
\put(2.5,3.5){\spos{}{\scriptstyle (0,0)}}
\put(0.5,-0.5){\spos{}{1}}
\put(1.5,-0.5){\spos{}{2}}
\put(2.5,-0.5){\spos{}{3}}
\put(3.5,-0.5){\spos{}{r}}
\put(-0.5,0.5){\spos{}{1}}
\put(-0.5,1.5){\spos{}{2}}
\put(-0.5,2.5){\spos{}{3}}
\put(-0.5,3.5){\spos{}{4}}
\put(-0.5,4.5){\spos{}{s}}
\end{picture}
}
\begin{picture}(5,5)
\put(2,6.){\spos{}{A_5\quad (k=4)}}
\multiput(0,0)(1,0){5}{\line(0,1){5}}
\multiput(0,0)(0,1){6}{\line(1,0){4}}
\put(0.5,0.5){\spos{}{\scriptstyle (4,2)}}
\put(1.5,0.5){\spos{}{\scriptstyle (0,2)}}
\put(2.5,0.5){\spos{}{\scriptstyle (4,0)}}
\put(3.5,0.5){\spos{}{\scriptstyle (4,0)}}
\put(0.5,1.5){\spos{}{\scriptstyle (1,1)}}
\put(1.5,1.5){\spos{}{\scriptstyle (3,1)}}
\put(2.5,1.5){\spos{}{\scriptstyle (1,3)}}
\put(3.5,1.5){\spos{}{\scriptstyle (1,3)}}
\put(0.5,2.5){\spos{}{\scriptstyle (2,2)}}
\put(1.5,2.5){\spos{}{\scriptstyle (2,2)}}
\put(2.5,2.5){\spos{}{\scriptstyle (2,0)}}
\put(3.5,2.5){\spos{}{\scriptstyle (2,0)}}
\put(0.5,3.5){\spos{}{\scriptstyle (3,1)}}
\put(1.5,3.5){\spos{}{\scriptstyle (1,1)}}
\put(2.5,3.5){\spos{}{\scriptstyle (3,3)}}
\put(3.5,3.5){\spos{}{\scriptstyle (3,3)}}
\put(0.5,4.5){\spos{}{\scriptstyle (0,2)}}
\put(1.5,4.5){\spos{}{\scriptstyle (4,2)}}
\put(2.5,4.5){\spos{}{\scriptstyle (0,0)}}
\put(3.5,4.5){\spos{}{\scriptstyle (0,0)}}
\put(0.5,-0.5){\spos{}{1}}
\put(1.5,-0.5){\spos{}{2}}
\put(2.5,-0.5){\spos{}{3}}
\put(3.5,-0.5){\spos{}{4}}
\put(4.5,-0.5){\spos{}{r}}
\put(-0.5,0.5){\spos{}{1}}
\put(-0.5,1.5){\spos{}{2}}
\put(-0.5,2.5){\spos{}{3}}
\put(-0.5,3.5){\spos{}{4}}
\put(-0.5,4.5){\spos{}{5}}
\put(-0.5,5.5){\spos{}{s}}
\end{picture}
} \vspace{.75cm} \caption{ The correspondence $(r,s)\mapsto (\ell,m)$
between the construction labels $(r,s)$ of the integrable boundary
conditions on the right and the parafermionic conformal fields
$(\ell,m)$ for the $A_{L}$ models with $L=3, 4, 5$ and number of faces
$N=1$ mod $2k$.}
        \label{tbl:corres1}
\end{table}

For a number of faces $N\neq 0$ mod $2k$, the left/right symmetry of
this correspondence is broken.  Keeping the same correspondence
between integrable boundary conditions $(r,s)$ and fields $(\ell,m)$
on the left hand side, the correspondence on the right hand side for a
number of faces $N=1$ mod $2k$ is simply given by the same table where
the columns are shifted cyclically to the left, the first column
becoming the last column after a top/bottom flip as shown in
Table~\ref{tbl:corres1} and so on for other mod properties. 
Therefore, for a number of faces $N=k$ mod $2k$, the correspondence
table is completely flipped about the middle row and we get the height
reversed correspondence.  This result is supported by our numerical
computations for $k=2,3,4$ summarised in Section~\ref{sec:Numerics}.

\begin{table}[htbp]
\nc{\spos}[2]{\makebox(0,0)[#1]{${#2}$}}
\nc{\sm}[1]{{\scriptstyle #1}}
\setlength{\unitlength}{9mm}
\begin{center}
\begin{picture}(7,3)
\multiput(0,0)(1,0){5}{\line(0,1){3}}
\multiput(0,0)(0,1){4}{\line(1,0){4}}
\put(4.02,0){\line(0,1){3}}
\put(.5,-1.25){\line(1,0){3}}
\multiput(.5,-1.25)(1,0){4}{\spos{}{\bullet}}
\put(0.5,0.5){\spos{}{\sm{(A,0)}}}
\put(1.5,0.5){\spos{}{\sm{(A,2)}}}
\put(2.5,0.5){\spos{}{\sm{(A,2)}}}
\put(3.5,0.5){\spos{}{\sm{(A,0)}}}
\put(0.5,1.5){\spos{}{\sm{(O,3)}}}
\put(1.5,1.5){\spos{}{\sm{(O,1)}}}
\put(2.5,1.5){\spos{}{\sm{(O,1)}}}
\put(3.5,1.5){\spos{}{\sm{(O,3)}}}
\put(0.5,2.5){\spos{}{\sm{(a,0)}}}
\put(1.5,2.5){\spos{}{\sm{(a,2)}}}
\put(2.5,2.5){\spos{}{\sm{(a,2)}}}
\put(3.5,2.5){\spos{}{\sm{(a,0)}}}
\put(0.5,-0.5){\spos{}{1}}
\put(1.5,-0.5){\spos{}{2}}
\put(2.5,-0.5){\spos{}{3}}
\put(3.5,-0.5){\spos{}{4}}
\put(-1,-0.5){\spos{}{r\in A_{4}}}
\end{picture}
\begin{picture}(5,3)
\multiput(0,0)(1,0){5}{\line(0,1){3}}
\multiput(0,0)(0,1){4}{\line(1,0){4}}
\put(4.02,0){\line(0,1){3}}
\put(-1.5,0.5){\line(0,1){1}}
\multiput(-1.5,0.5)(0,1){2}{\spos{}{\bullet}}
\put(-0.5,2.5){\spos{}{\bullet}}
\put(-2.5,2.5){\spos{}{\bullet}}
\put(-1.5,1.5){\line(1,1){1}}
\put(-1.5,1.5){\line(-1,1){1}}
\put(-.3,-1.25){\line(1,0){4.6}}
\multiput(.5,-1.25)(1,0){4}{\spos{}{\bullet}}
\multiput(-.05,-1.25)(1,0){5}{\spos{}{>}}
\put(0.5,0.5){\spos{}{\topped{(1,A)}{(4,A)}}}
\put(2.5,0.5){\spos{}{\topped{(3,A)}{(2,A)}}}
\put(1.5,1.5){\spos{}{\topped{(2,O)}{(3,O)}}}
\put(3.5,1.5){\spos{}{\topped{(1,O)}{(4,O)}}}
\put(0.5,2.5){\spos{}{\topped{(1,a)}{(4,a)}}}
\put(2.5,2.5){\spos{}{\topped{(3,a)}{(2,a)}}}
\put(0.5,-0.5){\spos{}{0}}
\put(1.5,-0.5){\spos{}{1}}
\put(2.5,-0.5){\spos{}{2}}
\put(3.5,-0.5){\spos{}{3}}
\put(5,-0.5){\spos{}{m\in \Bbb Z_{8}}}
\put(-1,0.5){\spos{}{A}}
\put(-1,1.5){\spos{}{O}}
\put(-0.5,3){\spos{}{C}}
\put(-2.5,3){\spos{}{B}}
\put(-1.5,3.5){\spos{}{a\in D_{4}}}
\end{picture}
\vspace{1cm}
\end{center}
	\caption{The correspondence $(r, a)\mapsto (a, m)$ between
	integrable boundary conditions $(r,a)$ and conformal fields
	$(a,m)$ for the $(D_{4},\Bbb Z_8)$ model with the number of
	faces $N=0$ mod 4.  Within these tables $a=B$ or $C$.  The
	minimal model Kac table symmetry reduces to the left/right
	flip in the $A_{4}\times D_{4}$ table.}
\label{tbl:corresD4}
\end{table}

\begin{table}[htbp]
        \centering
\nc{\spos}[2]{\makebox(0,0)[#1]{${#2}$}}
\nc{\sm}[1]{{\scriptstyle #1}}
\setlength{\unitlength}{9mm}
\begin{center}
\begin{picture}(4,3)
\multiput(0,0)(1,0){5}{\line(0,1){3}}
\multiput(0,0)(0,1){4}{\line(1,0){4}}
\put(4.02,0){\line(0,1){3}}
\put(-2.5,0.5){\line(0,1){1}}
\multiput(-2.5,0.5)(0,1){2}{\spos{}{\bullet}}
\put(-1.5,2.5){\spos{}{\bullet}}
\put(-3.5,2.5){\spos{}{\bullet}}
\put(-2.5,1.5){\line(1,1){1}}
\put(-2.5,1.5){\line(-1,1){1}}
\put(.5,-1.25){\line(1,0){3}}
\multiput(.5,-1.25)(1,0){4}{\spos{}{\bullet}}
\put(0.5,0.5){\spos{}{\sm{(A,2)}}}
\put(1.5,0.5){\spos{}{\sm{(A,2)}}}
\put(2.5,0.5){\spos{}{\sm{(A,0)}}}
\put(3.5,0.5){\spos{}{\sm{(A,0)}}}
\put(0.5,1.5){\spos{}{\sm{(O,1)}}}
\put(1.5,1.5){\spos{}{\sm{(O,1)}}}
\put(2.5,1.5){\spos{}{\sm{(O,3)}}}
\put(3.5,1.5){\spos{}{\sm{(O,3)}}}
\put(0.5,2.5){\spos{}{\sm{(a,2)}}}
\put(1.5,2.5){\spos{}{\sm{(a,2)}}}
\put(2.5,2.5){\spos{}{\sm{(a,0)}}}
\put(3.5,2.5){\spos{}{\sm{(a^{*},0)}}}
\put(0.5,-0.5){\spos{}{1}}
\put(1.5,-0.5){\spos{}{2}}
\put(2.5,-0.5){\spos{}{3}}
\put(3.5,-0.5){\spos{}{4}}
\put(5,-0.5){\spos{}{r\in A_{4}}}
\put(-2,0.5){\spos{}{A}}
\put(-2,1.5){\spos{}{O}}
\put(-1.5,3){\spos{}{C}}
\put(-3.5,3){\spos{}{B}}
\put(-2.5,2.5){\spos{}{a\!\in\! D_{4}}}
\end{picture}
\end{center}
	\vspace{1cm} \caption{The correspondence $(r, a)\mapsto (a,
	m)$ between integrable boundary conditions $(r,a)$ and
	conformal fields $(a,m)$ for the $(D_{4},\Bbb Z_8)$ model with
	the number of faces $N=1$ mod 4.  Note that $B^{*}=C$ and
	$C^{*}=B$.}
      \label{tbl:corresD41}
\end{table}

This correspondence can be carried over to the $D$ and $E$ cases.  For
$N=0$ mod $k$, the result is again given by the trivial identification
of $A_{k}$ with $\Bbb Z_{k}$ for the even sublattice and its flip for
the odd sublattice.  The cylinder partition functions in this case are
obtained from the $A$ cases by intertwiners~\cite{BehrendP}
\begin{equation}
       Z_{r_{1}a_{1}|r_{2}a_{2}}(q)=\sum_{s=1}^{L}V_{s\; a_{1}}{}^{a_{2}}
       Z^{A_{L}}_{r_{1}1|r_{2}s}(q)
       \label{eq:fusion}
\end{equation}
Comparing our numerical results with Table~\ref{tbl:cylinderExtD4}
allows us to obtain the correspondence given in
Table~\ref{tbl:corresD4} for the $(D_{4},\Bbb Z_8)$ case with the
number of faces $N=0$ mod 4:
\begin{equation}
    \begin{array}{ccl}
         A_{4}\times D_{4} & \to &\phantom{123} D_{4}\times \Bbb Z_{8} \\
        (r,a)& \mapsto & \cases{(a, r-1)&\text{ if $r+a$ is odd,}\\
     (a, 4-r)&\text{ if $r+a$ is even.}}
    \end{array}
     \label{eq:correspondenceD4}
\end{equation}

For example, the partition function for the boundary conditions
$(r,a)=(1,O)$ on the left and right is:
\begin{eqnarray}
     Z^{D_{4}}_{1O|1O}(q) & = &
     \sum_{s=1}^{5}V_{s\;O}{}^{O}Z^{A_{5}}_{11|1s}(q)
           \\
      & = & Z^{A_{5}}_{11|11}(q)+2Z^{A_{5}}_{11|13}(q)+Z^{A_{5}}_{11|15}(q)
     \label{eq:ZD4}
      = \chi_{(0,0)}(q)+2\chi_{(2,0)}(q)+\chi_{(4,0)}(q).\notag
\end{eqnarray}
This result agrees with the alternative way of computing this
partition function using the
correspondence~(\ref{eq:correspondenceD4}) and the extended fusion
algebra described in Table~\ref{tbl:cylinderExtD4}:
\begin{eqnarray}
     Z^{D_{4}}_{1O|1O}(q) & = & Z^{D_{4}}_{(O,3)|(O,3)}(q)
           \\
      & = & \hat\chi_{(A,0)}(q)+ \hat\chi_{(B,0)}(q)+ \hat\chi_{(C,0)}(q)
     \label{eq:ZExtD4}
     =\chi_{(0,0)}(q)+\chi_{(4,0)}(q)+\chi_{(2,0)}(q)+\chi_{(2,0)}(q).\notag
\end{eqnarray}
More generally, the correspondence in the $D$ and $E$ cases relate
(\ref{eq:block}) and~(\ref{eq:fusion}) as two compatible ways of
computing partition functions in terms of $A_{L}$ partition functions.

As in the $A_{L}$ case, the table for $N=1$ mod 4 is obtained by a
shifting procedure on the table, namely, shift the columns of the
table to the left and interchange the $B$ and $C$ entries in the last
column (see Table~\ref{tbl:corresD41}).

Using the intertwiner definition (\ref{eq:fusion}) and the
correspondence formula~(\ref{eq:correspondence}), one finds that this
result holds for any $D_{\frac{k}{2}+2}$ and for the exceptional cases
$E_{6}$, $E_{7}$ and $E_{8}$:
\begin{equation}
    (a,m)=\cases{(a, r-1)&\text{ if $r+a$ is odd,}\\
     (a, k-r)&\text{ if $r+a$ is even}}
     \label{eq:correspondenceDE}
\end{equation}
where $a\in G$, $r\in A_k$ and $m\in{\Bbb Z}_{2k}$.  This formula also
holds for $G=A_{k+1}$ provided the ranges of $r$ and $s$ are adjusted
to begin at 1 rather than 0.  Here the labelling of the nodes of the
$D$ and exceptional $E$ graphs is as follows: \vspace{-.2in}
\begin{center}
\nc{\spos}[2]{\makebox(0,0)[#1]{{$\scriptstyle #2$}}}
\nc{\sm}[1]{{\scriptstyle #1}}
\setlength{\unitlength}{6mm}
     \begin{picture}(6.5,3)
	\put(-1,0){\spos{}{D_{\frac k2 +2}:}}
\put(0,0){\line(1,0){1.5}}
\put(2.5,0){\line(1,0){.5}}
\multiput(0,0)(1,0){2}{\spos{}{\bullet}}
\put(3,0){\spos{}{\bullet}}
\put(3,0){\line(1,1){1}}
\put(3,0){\line(1,-1){1}}
\put(4,1){\spos{}{\bullet}}
\put(4,-1){\spos{}{\bullet}}
\put(0,0.5){\spos{}{1}}
\put(1,0.5){\spos{}{2}}
\put(2,0){\spos{}{\cdots}}
\put(3,0.5){\spos{}{\frac k2}}
\put(4,1.5){\spos{}{\frac k2+1}}
\put(4,-1.5){\spos{}{{(\frac k2+1)}^{*}}}
\end{picture}
\end{center}

\vspace{.4in}
\mbox{}\hspace{.15in}
\vbox{
\nc{\spos}[2]{\makebox(0,0)[#1]{{$\scriptstyle #2$}}}
\nc{\sm}[1]{{\scriptstyle #1}}
\setlength{\unitlength}{6mm}
\begin{center}
     \begin{picture}(6,1)
	\put(-.75,0){\spos{}{E_{6}:}}
\put(0,0){\line(1,0){4}}
\multiput(0,0)(1,0){5}{\spos{}{\bullet}}
\put(2,0){\line(0,1){1}}
\put(2,1){\spos{}{\bullet}}
\put(0,0.5){\spos{}{1}}
\put(1,0.5){\spos{}{2}}
\put(2,-0.5){\spos{}{3}}
\put(3,0.5){\spos{}{4}}
\put(4,0.5){\spos{}{5}}
\put(2,1.5){\spos{}{6}}
\end{picture}
     \begin{picture}(7,1)
	\put(-.75,0){\spos{}{E_{7}:}}
\put(0,0){\line(1,0){5}}
\multiput(0,0)(1,0){6}{\spos{}{\bullet}}
\put(3,0){\line(0,1){1}}
\put(3,1){\spos{}{\bullet}}
\put(0,0.5){\spos{}{1}}
\put(1,0.5){\spos{}{2}}
\put(2,0.5){\spos{}{3}}
\put(3,-0.5){\spos{}{4}}
\put(4,0.5){\spos{}{5}}
\put(5,0.5){\spos{}{6}}
\put(3,1.5){\spos{}{7}}
\end{picture}
     \begin{picture}(8,1)
	\put(-.75,0){\spos{}{E_{8}:}}
\put(0,0){\line(1,0){6}}
\multiput(0,0)(1,0){7}{\spos{}{\bullet}}
\put(4,0){\line(0,1){1}}
\put(4,1){\spos{}{\bullet}}
\put(0,0.5){\spos{}{1}}
\put(1,0.5){\spos{}{2}}
\put(2,0.5){\spos{}{3}}
\put(3,0.5){\spos{}{4}}
\put(4,-0.5){\spos{}{5}}
\put(5,0.5){\spos{}{6}}
\put(6,0.5){\spos{}{7}}
\put(4,1.5){\spos{}{8}}
\end{picture}
\end{center}
}

\section{Numerical Conformal Spectra} \label{sec:Numerics}

In this section we describe our numerical analysis which confirm the
results presented in Section~\ref{sec:Identify}.  As explained in the
previous section, the spectra for the $D$ and $E$ models are related
to the spectra of the $A$ models by intertwiners (\ref{eq:fusion}) so
we need only do numerics on the $A$ parafermion models.  The
Yang-Baxter equation, boundary Yang-Baxter equation, reflection
symmetry, inversion relation, crossing symmetry, height reversal
symmetry and commutation of double row transfer matrices for various
number of faces and all boundary conditions were checked numerically. 
The numerics were carried out for finite size row transfer matrices
and extrapolated to large sizes.  The spectra were obtained by
numerical diagonalization of the row transfer matrices.  For the
$A_{L}$ lattice models the maximum number of faces $N$ in a row that
we found manageable due to machine limitations was $22$, $18$ and $16$
respectively for $L=3, 4, 5$.

We first studied the boundary conditions which from (\ref{singchar})
should yield single characters, such as $(r,1|1,1)$ or $(1,s|1,1)$
with the number of faces $N=0$ mod $2k$.  We computed numerically the
double row transfer matrices for these boundary conditions for
increasing $N$ and numerically diagonalized them to obtain the
spectra.  Allowing for the contribution of the bulk and boundary free
energies, the central charge and the geometric factor, we fit the data
corresponding to the largest eigenvalues of these transfer matrices,
in negative integers powers of $N$ to extract the conformal weight
$\Delta$.  This value was compared with the entries in the parafermion
Kac table to determine (up to conjugacy) the primary field associated
with the given boundary condition.

In the $A_{5}$ case the level is $k=4$ and machine limitations prevent
us from dealing with more than $N=16$ faces.  Consequently, this
direct method would give us only two numbers, for $N=8$ and $N=16$
faces respectively.  However, some symmetries can be used to obtain
more data for extrapolation to large $N$ by merging sequences for
related boundary conditions and other mod properties as we now
explain.  Firstly, height reversal symmetry allows us to merge the
sequence $(r,s|1,1)$ for $N=0$ mod $2k$ with the sequence $(r,s|1,L)$
for $N=k$ mod $2k$.  More generally, other sequences converge to the
same character, namely $(r,s|m,1)$ for $N=2k-m$ mod $2k$ and
$(r,s|m,L)$ for $N=k-m$ mod $2k$.  Indeed, these boundary conditions
are all compatible with the sawtooth shaped ground state.  We
discovered that this shift and flip procedure not only applies for the
vacuum but for any boundary condition (see Table~\ref{tbl:corres1}). 
This method of interleaving sequences allowed us to improve
dramatically the accuracy of our numerics.  A typical extrapolated
numerical estimate of a conformal weight agrees to 5 or 6 digit
accuracy with the exact entry in the parafermion Kac table.

In addition to producing the correct conformal weight $\Delta$, the
numerical spectra should, in the case of a single character, also
reproduce the correct degeneracies of the associated string function
$\chi_{(\ell ,m)}$.  This means that if the eigenvalues
$\Lambda_n^{(N)}$ of the double row transfer matrix with $N$ faces are
placed in decreasing order with $n=0,1,2,\ldots$ then
\begin{equation}
\Lambda_n^{(N)}=
\Lambda_0^{(N)}\exp\Big(-\frac{2\pi\sin\vartheta}N\;{a_n}+o\big(\frac 1
N\big)\Big)
\end{equation}
where $a_n$ is an increasing integer sequence encoding the
degeneracies in the character in terms of the modular parameter
$q=\exp(-\frac{2\pi}N\sin\vartheta)$.  The leading corrections in the
$o(\frac 1 N)$ term are negative integer powers of $N$ so a plot of
the normalized $\log\mbox{}$ value against $1/N$ becomes a straight
line near zero and a polynomial fit in $1/N$ gives a good
extrapolation to $N=\infty$.

\begin{figure}[htbp]
\begin{picture}(10,11)
\epsfig{file=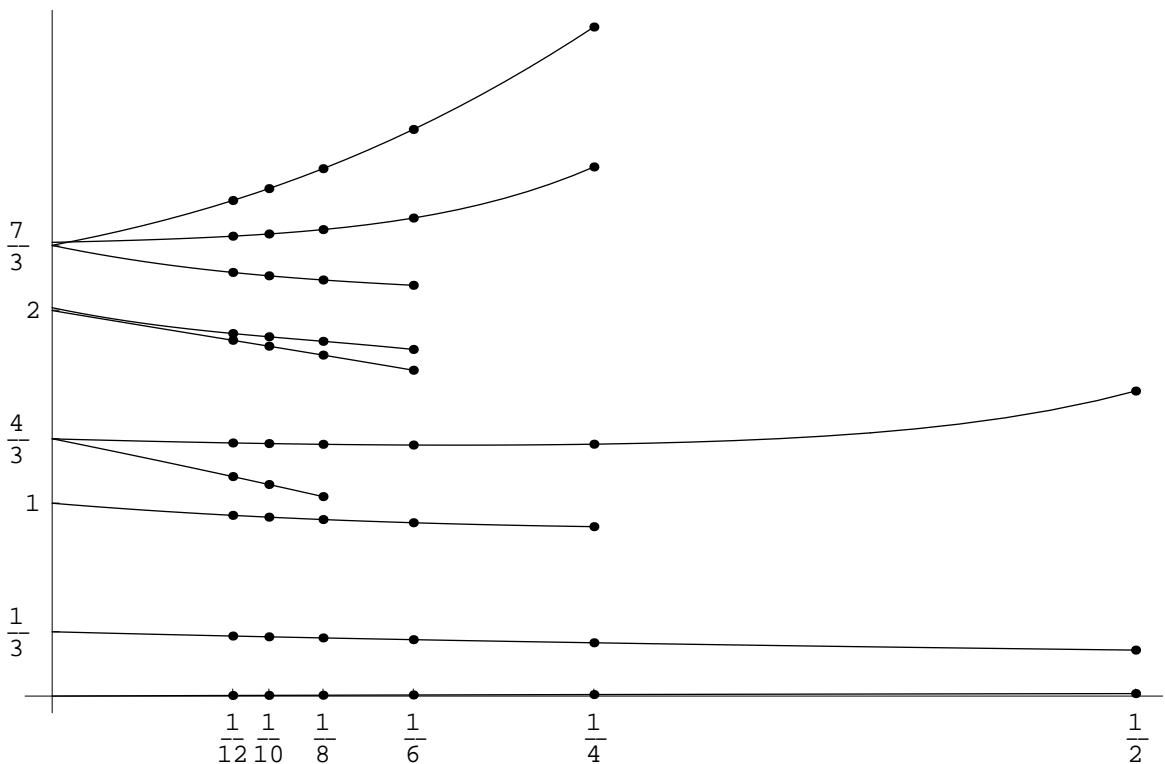}
\end{picture}
\caption{The extrapolated sequences corresponding to the first ten
energy levels of the double row transfer matrix for the $A_5$
parafermion model with $(1,3|1,3)$ boundary conditions.  The
horizontal axis is $1/N$ where $N$ is the number of faces.}
\label{fig:character}
\end{figure}
\begin{table}
\begin{tabular}{c|cccccccccc}
     Energy & 1 & 2 & 3 & 4 & 5 & 6 & 7 & 8 & 9 & 10 \\
     \hline
    Numerical & $\scriptstyle{1.04\cdot 10^{-6}}$ &
    $\scriptstyle{0.33331}$ & $\scriptstyle{0.999502}$ &
    $\scriptstyle{1.33363}$ & $\scriptstyle{1.33304}$ &
    $\scriptstyle{1.99795}$ & $\scriptstyle{2.01268}$ &
    $\scriptstyle{2.35139}$ & $\scriptstyle{2.33541}$ &
    $\scriptstyle{2.3362}$\\
     Exact  & $0$ & $\frac13 $ & $1$ & $\frac43$ & $\frac43$ & $2$ & $2$
     & $\frac73$ & $\frac73$ & $\frac73$
\end{tabular}
\caption{Comparison of the exact values and numerical estimates of the
first ten energy levels for the $A_5$ parafermion model with
$(1,3|1,3)$ boundary conditions.  The cylinder partition function is
$Z_{13|13}(q)=Z_{(2,0)|(2,0)}(q)=\chi_{(0,0)}(q)+\chi_{(2,0)}(q)+\chi_
{(4,0)}(q)$.}
\label{tab:character}
\end{table}

After investigating all the boundary conditions leading to a single
character we turned to the other boundary conditions and compared the
numerical spectra with the predictions of the parafermion fusion
algebra.  In these cases the cylinder partition functions are sums of
characters.  As a typical example, consider the $(A_{5},\Bbb Z_{8})$
parafermion model with $(1,3|1,3)$ boundary conditions and $N=0$ mod
$8$.  According to Table~\ref{tbl:corres}, the normalized partition
function should converge to the sum of characters:
\begin{eqnarray}
      Z_{13|13}(q)  & = & Z_{(2,0)|(2,0)}(q)=
\chi_{(0,0)}(q)+\chi_{(2,0)}(q)+\chi_{(4,0)}(q)
        \label{eq:Z1313} \\
   & =&q^{-1/24}(1+q^{1/3}+q+2q^{4/3}+2q^2+3q^{7/3}+5q^3
     +o(q^3))\notag
   \end{eqnarray}
so the energy levels and their degeneracies should be given by the
sequence
\begin{equation}
\{a_n\}=\{0,\frac 13,1,\frac 43,\frac 43,2,2,\frac 73,\frac 73,\frac
73,3\ldots\}
\end{equation}
Table~\ref{tab:character} and Figure~\ref{fig:character} show a
comparison of the extrapolated sequences and the exact values.  The
accuracy in other cases is at least as good as this case.  As can be
seen in Figure~\ref{fig:character}, in extrapolating the sequences,
the eigenvalues cannot simply be ordered according to their values at
a given $N$.  However, a given eigenvalue can be uniquely identified
and tracked for each $N$ by examining its pattern of zeros in the
complex $u$-plane.

The $D$ and $E$ parafermion models can be studied numerically by the
same direct numerical diagonalization techniques as the $A$-type
models.  However, in these cases, it is better to use the intertwiner
relations (\ref{eq:fusion}) to relate~\cite{PZ93} the spectra of the
$D$ or $E$ model to the spectra of various sectors (or boundary
conditions) of the associated $A$ model with the same Coxeter number. 
In this way, for example, the spectra of the $D_{4}$ model with given
boundary conditions is related to sectors of the $A_{5}$ model. 
Although it is not necessary to directly diagonalize the double row
transfer matrices of the $D$ and $E$ parafermion models we did
diagonalize some cases as checks on the numerics.

\section{Discussion}				\label{sec:Discussion}

In this paper we have shown how to associate integrable lattice
boundary conditions to each conformal boundary condition $(a,m)$ of
the $A$-$D$-$E$ $\hat{s\ell}(2)$ parafermion models.  Moreover, we
have shown how the conformal labels $(a,m)$ of the string functions
are related to the labels $(r,a)$ that naturally arise in the
construction of the integrable boundary conditions.  In contrast to
the unitary minimals models, we observe (\ref{chiral}) that the
symmetry between the left and right boundaries is broken for the
parafermion models due to the presence of an intrinsic cyclic
chirality.  We note also that there does not exist a gauge in which
the local Boltzmann weights of the parafermion lattice models are all
positive.  Since the local weights should represent probabilities,
this means that these models are not well defined as lattice models in
statistical mechanics.  Nevertheless, as is also the case for the
non-unitary minimal models, these models yield well-defined conformal
field theories in the continuum scaling limit.

In this work we have focussed on the principal parafermion models. 
It would be of interest to extend our results to the non-principal
models.  We expect these models to correspond to fused
versions~\cite{KP92, BPO'B96} of the lattice models studied here.
\section*{Acknowledgements} 			\label{sec:Acknowldegments}
This research is supported by the Australian Research Council. We thank 
Roger Behrend for his careful reading and comments.

\bibliographystyle{UNSRT} \bibliography{ZnParafermions}

\end{document}